\documentclass[twocolumn]{aastex62}

\usepackage{amsmath}
\received{January 1, 2018}
\revised{January 7, 2018}
\accepted{\today}
\submitjournal{ApJ}

\shorttitle{Non-standard modeling of KIC 11145123}
\shortauthors{Hatta et al.}
\begin{document}

\title{Non-standard modeling of a possible blue straggler star, KIC 11145123}

\correspondingauthor{Yoshiki Hatta}
\email{yoshiki.hatta@grad.nao.ac.jp}

%\correspondingauthor{August Muench}
%\email{greg.schwarz@aas.org, gus.muench@aas.org}

\author[0000-0003-0747-8835]{Yoshiki Hatta}
\affiliation{Department of Astronomical Science, School of Physical Sciences, SOKENDAI\\
2-21-1 Osawa, Mitaka, Tokyo 181-8588, Japan}
\affiliation{National Astronomical Observatory of Japan \\
2-21-1 Osawa, Mitaka, Tokyo 181-8588, Japan}

\author[0000-0001-6583-2594]{Takashi Sekii}
\affiliation{Department of Astronomical Science, School of Physical Sciences, SOKENDAI\\
2-21-1 Osawa, Mitaka, Tokyo 181-8588, Japan}
\affiliation{National Astronomical Observatory of Japan \\
2-21-1 Osawa, Mitaka, Tokyo 181-8588, Japan}

%\collaboration{(AAS Journals Data Scientists collaboration)}

\author[0000-0001-9430-001X]{Masao Takata}
\affiliation{Department of Astronomy, School of Science, The University of Tokyo\\
Bunkyou-ku, Tokyo 113-0033, Japan}

\author[0000-0001-9405-5552]{Othman Benomar}
\affiliation{Department of Astronomical Science, School of Physical Sciences, SOKENDAI\\
2-21-1 Osawa, Mitaka, Tokyo 181-8588, Japan}
\affiliation{National Astronomical Observatory of Japan \\
2-21-1 Osawa, Mitaka, Tokyo 181-8588, Japan}

%\affiliation{Jeremiah Horrocks Institute, University of Central Lancashire\\
%Preston PR1 2HE, UK}



\begin{abstract}
Non-standard modeling of KIC 11145123, a possible blue straggler star, has been asteroseismically carried out 
based on a scheme to compute stellar models 
with the chemical compositions in their envelopes arbitrarily modified, 
mimicking effects of some interactions with other stars through which blue straggler stars are thought to be born. 
%This is the first time such asteroseismic non-standard modeling has been performed 
%for a blue straggler star. 
We have %succeeded in constructing a non-standard model of the star with the following 
constructed a non-standard model of the star with the following 
parameters: $M=1.36M_{\odot}$, $Y_{\mathrm{init}}=0.26$, 
$Z_{\mathrm{init}}=0.002$, and $f_{\mathrm{ovs}}=0.027$, %and $ \mathrm{Age}=2.169 \times 10^{9}$ years old, 
where $f_{\mathrm{ovs}}$ is the extent of overshooting described as an exponentially decaying diffusive process. 
The modification is down to the depth of $r/R\sim0.6$ and the extent $\Delta X$, % \sim 0.06$ 
which is a difference in surface hydrogen abundance between the envelope-modified and unmodified models, % at the surface, 
is $0.06$. 
%corresponds to a hydrogen abundance difference of $6\%$ 
%between the unmodified and envelope-modified models at the surface. 
%($\Delta X$ is a difference in hydrogen abundance between the candidate model and the modified model) at the surface. 
The residuals between the model and the observed frequencies are comparable with those for the previous models
computed assuming standard single-star evolution, suggesting that 
it is possible that the star was born with an relatively ordinary initial helium abundance of $\sim0.26$ 
compared with that of the previous models ($\sim0.30$--$0.40$), 
then experienced some modification of the chemical compositions, and gained helium in the envelope. 
Detailed analyses of the non-standard model have implied that the elemental diffusion 
in the deep radiative region of the star might be much weaker than that assumed in 
current stellar evolutionary calculations; we need some extra mechanisms inside the star, 
%
%there 
%might be a relation between the chemical composition gradient in the deep radiative region and the 
%previously inferred rotational velocity shear between the convective core and the radiative region above, 
rendering the star a much more intriguing target to be further investigated. 
\end{abstract}

 %; Stellar diffusion (1593)}


\section{Introduction} \label{sec:intro}
Space-borne missions such as %MOST \citep{MOST}, 
%CoRoT \citep{CoRoT}, 
Kepler \citep{Kepler} and TESS \citep{TESS} have enabled us to conduct 
extremely precise measurements of stellar variabilities ($\sim$ $10^{-5}$ magnitude) 
with short cadences ($\sim$ minutes) and long durations ($\sim$ months to years). 
In particular, asteroseismology, a branch of stellar physics in which we probe the interiors of stars 
based on the measurements of stellar oscillations \citep[e.g.][]{Aerts_text}, 
has been greatly benefitting from such high-quality observations carried out by the modern spacecrafts, 
and we are now able to perform asteroseismic analyses 
to identify evolutionary stages \citep[e.g.][]{Bedding2011, Chaplin2013} 
and to investigate internal structures \citep[e.g.][]{Kosovichev2020} and dynamics \citep[e.g.][]{Aerts2019} of stars in detail, 
shedding new light on the understanding of stellar interiors from the observational point of view. 
%which the other branches have not been able to do. 

Among stars thus asteroseismically analyzed so far, KIC 11145123, which is one of the Kepler targets and 
photometrically categorized as a main-sequence A-type star \citep{Huber2014}, 
outstands in terms of its well-resolved frequency splittings for p, g, and mixed modes, whose restoring forces are 
pressure, buoyancy (gravity), and either of them (depending on where the modes are established), %propagate), 
respectively \citep[see, e.g.,][]{Unno_text}. 
The well-determined frequencies and frequency splittings have allowed us 
to perform a number of detailed asteroseismic analyses of the star 
inferring, for instance, the internal rotation profile \citep{Kurtz2014, Hatta2019}, the asphericity \citep{Gizon2016}, 
and the evolutionary stage of the star \citep{Kurtz2014, Takada_Hidai2017}. 
It should be instructive to note that the number of main-sequence stars in Kepler targets which exhibit well-resolved 
frequency splittings for both p and g modes 
is just three \citep{Saio2015, Schmid_Aerts2016} including KIC 11145123; the star therefore could be an important 
testbed which may be helpful for putting observational constraints on theoretical studies of the stellar interiors. 

There is, however, an issue on previous equilibrium models 
and the thus identified evolutionary stage of the star \citep{Kurtz2014, Takada_Hidai2017}. 
\citet{Kurtz2014} is the first to construct a 1-dimensional model of the star 
to reproduce the observed frequencies via Modules for Experiments in Stellar Astrophysics \citep[MESA; version 4298,][]{Paxton2013} 
% MESA (refs), 
and they noticed that higher 
initial helium abundance of $\sim0.3$--$0.4$ is much favored. % to reproduce the observed p-mode frequencies. 
But such higher initial helium abundance is difficult to explain for a simple single star 
considering the ordinary stellar evolution, %and the expected helium abundance of $\sim0.25$ in the earliest Universe 
%realized by the Big Bang nucleo-synthesis \citep{Cyburt2016}, 
and thus, we need some mechanisms to render the helium abundance of the star so high. 

One simple scenario is that the star experienced some interactions with other stars such as mass accretion, 
stellar merger, or stellar collision, and it has obtained extra helium from the outside. 
%, and thus, \citet{Kurtz2014} 
%pointed out the possibility that the star is a blue straggler star which is thought to be born via some interactions with the other stars (see 
%Subsection \ref{1-1-2}.) 
Interestingly, the inferred internal rotation profile of the star, where 
the outer envelope is rotating slightly faster than the deep radiative region \citep{Kurtz2014, Hatta2019}, 
is also pointing toward the same scenario; we need some mechanisms of angular momentum transfer, 
probably from the outside, to realize the inferred 
rotational profile, and interacting with other stars can be a straightforward explanation. 
Based on the suggestion that the star has experienced some interactions during the evolution, 
in addition to the relatively lower initial metallicity of $\sim 0.010$ determined by asteroseismic modeling of the star, 
\citet{Kurtz2014} finally pointed out that the star could be a blue straggler star \citep{Sandage1953}, 
which is a kind of stars thought to be born via binary interactions or stellar collisions \citep{Boffin2015} 
and appears somehow rejuvenated compared with other stars at the same age. 
%and a suggested detection of the fast-convective-core rotation (ref and tackled in detail in ref in prep.). 
%Especially, we have not succeeded in constructing a 1-dimensional equilibrium model of the star... 
%Kurtz... 
%initial helium abundance... too high. 
%env. rotating slightly faster than the deep radiative region, AMT from the outside? 
%implying the star to be a bss. 

%BSS is a ... cite from parts of my dissertation! 
%and also mention that there are a kind of peculiar single stars with high helium abundances. 

To better comprehend properties of a star from a different perspective, 
\citet{Takada_Hidai2017} have conducted the spectroscopic observation of the star with Subaru/HDS. 
They firstly found that the star has a sub-solar iron content 
$[\mathrm{Fe}/\mathrm{H}] \sim -0.7$ which approximately corresponds to $Z\sim 0.003$. 
Based on the low metallicity, which was below the parameter range surveyed by \citet{Kurtz2014}, 
they have modeled the star assuming single-star evolution, 
%resulting in the higher initial helium abundance as that of \citet{Kurtz2014}; 
resulting in the initial helium abundance as high as that of \citet{Kurtz2014}; 
the situation has not improved. 
%
%They firstly found that 
%their best model of the star, asteroseismically constructed 
%based on the low metallicity ($Z \sim 0.003$) spectroscopically inferred, 
%prefers higher initial helium abundance as that of \citet{Kurtz2014}, 
%which is still too high for an ordinary single star. 
Importantly, however, based on the abundance pattern of the star, 
they found the star to be spectroscopically a blue straggler star as suggested by \citet{Kurtz2014}. %
%Since blue straggler stars are thought to have experienced binary interactions or stellar collisions, 
The star has probably experienced some interactions with other stars, which could be a reason 
for the high initial helium abundance previously deduced via modeling the star assuming single-star evolution; 
there is room for modeling the star in a non-standard manner where we take such interactions with other stars during the evolution 
into account. 

%TH, still high initial helium abundance even with the lower initial Z. 
%BSS! spectroscopically. 
%Let us model the star in a non-standard manner, taking some interactions into account as well. 

%Another possible 
%It is worth mentioning that there have been no attempts to carry out non-standard modeling of blue straggler stars 
%focusing on reproducing observed eigenfrequencies...
%the attempt  modeling of BSS... not established...
%In addition, ...
%Why have the high initial helium abundances been suggested for the star in the previous studies? 

%The other possibility is that the star was originally born as a star with an ordinary initial helium abundance, and somehow obtain the helium from the outside, leading the star to be a blue straggler with high helium abundance in its envelope. 
%Still, we have not had a clear answer to that point so far. (and the non-standard modeling helps us answer the question)

The primary goal of this paper is to construct a 1-dimensional non-standard model of a possible blue straggler star, KIC 11145123, 
%qualitatively 
taking into account effects of some interactions with other stars. % into account. 
 %, for instance, mass accretion or stellar collision.  
%envelope modification caused by mass accretion or stellar collision. 
Though there have been numerous attempts of non-standard modeling of blue straggler stars \citep[e.g.][]{Brogaard2018}, 
this is the first time such non-standard modeling of blue straggler stars has been carried out asteroseismically. 
In addition, the non-standard model would be of a great value 
since we can further perform detailed analyses such as rotation inversion of the star (Hatta et al. in prep.) 
%Note that we do not intend to 

%It should be noticed that 
As a first step toward the goal described in the previous paragraph, 
we specifically concentrate on effects of modifications of chemical compositions, 
namely, helium enhancements, in the envelope 
caused by whatever can lead to such envelope modification. 
In other words, we do not consider detailed physics concerning specific processes such as mass accretion or stellar merger, 
and what we do consider is just the resultant effects of these processes on stellar structure. 
%This is partly because no companion has been confirmed for the star based on the spectroscopic analyses of \citet{Takada_Hidai2017}; 
%we cannot determine an exact process (for instance, mass accretion or stellar merger) via which 
%it is not evident what an exact process (for instance, mass accretion or stellar merger) is, via which 
%the star has obtained helium and become a blue straggler star. 
Since there has been no scheme to realize such computations to our knowledge, we would like to 
develop the scheme in this study, which is actually the minor (though necessary) goal in this study. 

It is also worth mentioning that there do exist single stars with high initial helium abundances of $0.30$--$0.40$ 
which are thought to be born in such high helium environments contaminated by 
stellar winds from already existing asymptotic giant branch stars. 
This multiple main-sequence phenomenon \citep{Bastian2018} has been recently observed for some globular clusters. 
Although KIC 11145123 is currently not considered as such a peculiar single star 
based on the chemical abundance pattern and the kinematics determined by \citet{Takada_Hidai2017}, 
%Although KIC 11145123 is a field star based on its stellar kinematics determined by \citet{Takada_Hidai2017} 
%and it is thus currently not in the high helium environments, 
%we can also test whether the star is the kind of peculiar single star kicked out of some globular cluster or not, 
%by constructing a non-standard model of the star and 
%comparing the non-standard model with the previous (standard) ones. 
we can further deepen the discussion about the evolutionary history of the star from a different perspective 
by constructing a non-standard model of the star, comparing the non-standard model with the previous (standard) models, 
and finally assessing whether the initial helium abundance of the star was really high or not. 

The structure of this paper is as follows. 
After we briefly explain general setups for computing stellar models and frequencies in Section \ref{sec:2.5}, 
%In Section \ref{sec:2}, 
a scheme of %the nonstandard modeling and 
calculating 1-dimensional stellar models, whose chemical compositions are arbitrarily modified, and 
examples of envelope-modified models computed via the scheme are presented in Section \ref{sec:2}. 
Because the scheme is newly developed, we demonstrate the concept and the mathematical formulations in detail. 
In Section \ref{sec:3}, the non-standard model of the star is obtained %following the procedures 
by applying the scheme explained in Section \ref{sec:2}. 
Section \ref{sec:4} is devoted to discussions on the evolutionary stage, the internal structure, and a possible relation 
between the internal structure and the rotational velocity shear, the latter of which has been inferred by \citet{Hatta2019}. 
We conclude in Section \ref{sec:5}. 
 
\section{Numerical codes} \label{sec:2.5}
%We have some notes on community codes we have used in this study 
%for computing stellar models and eigenfrequencies. 
Stellar models are computed via MESA \citep[version 9793,][]{Paxton2015}. 
Neither rotation nor magnetic fields is assumed in the computation. 
The OPAL tables are used for both the equation of state and the opacity, and 
the nuclear reaction rate is obtained by interpolation based on a built-in table in MESA, called `basic.net'. 

As mixing processes during the evolution, convection, convective overshooting, and elemental diffusion are activated. 
The free parameter in the Mixing Length Theory $\alpha_{\mathrm{MLT}}$ is fixed to be $1.7$ following \citet{Kurtz2014}. 
Convective overshooting is implemented as a diffusive process whose diffusion coefficient follows an exponential decay 
above the convective boundary \citep{Herwig2000}. 
The effect of elemental diffusion is incorporated into the evolutionary calculation 
by solving so-called Burger's equation which is based on Boltzman's equation in kinematics \citep{Burger1969}. 
In our calculations, radiative levitation is not included. 
Note that though the scheme for elemental diffusion is 
considered to be working well for models with $\sim 1M_{\odot}$, it has been long pointed out that 
the scheme often overestimates the diffusion velocity of helium in the outermost layer of models with $> 1.3M_{\odot}$, 
sometimes leading to the depletion of helium there, 
which has not been observationally confirmed \citep{Morel2002}. 
To avoid such depletion of helium 
in the outermost layer of the models, %, which can be caused by neglecting radiative levitation, a 
a special scheme $\mathrm{radiation} \_ \mathrm{turbulence} \_ \mathrm{coeff}$ \citep{Morel2002} is 
turned on and set to be unity in our computations. 
For more details, see a series of papers about the code \citep{Paxton2011, Paxton2013, Paxton2015, Paxton2018, Paxton2019}. 

Eigenfrequencies of stellar models are computed via linear oscillation code GYRE \citep{Townsend2013}. 
The effects of rotation, magnetic fields, 
asphericity, and nonadiabaticity on the eigenoscillations of a certain model are not taken into account; 
these effects are considered to be sufficiently small for the star % and there is no observational hint for the effects 
\citep[e.g.][]{Kurtz2014}. 
 
\section{Method} \label{sec:2}
In this section, a scheme %for non-standard modeling, 
of constructing 
%which allows us to construct 
1-dimensional stellar models whose chemical compositions %in the envelopes 
are arbitrarily modified is presented. % which is formulated %based on 
The scheme is formulated similarly to the linear adiabatic radial oscillation of stars. %, 
%\edit1{the theory of linear adiabatic perturbations}, 
%the linear adiabatic radial oscillation equation, 
%to be applied to non-standard modeling of the star in Section \ref{sec:3}, 
%is presented. 
%The scheme can be shortly summarized as follows. 
Since the scheme is new (though the scheme itself is simple enough for us to incorporate it 
with the existing codes such as MESA), 
we demonstrate the concept (Section \ref{sec:2-1}), the mathematical formulations (Section \ref{sec:2-2}), 
and examples of modified models and the frequencies computed via the scheme (Section \ref{sec:2-3}) in detail. 
%We finally point out a few advantages of the developed scheme by 
%comparing with a rather simple scheme (Section \ref{sec:2-4}). 
We see the applications of the scheme to the non-standard modeling of KIC 11145123 in Section \ref{sec:3}. 

Before going into details, we have two specific notes as follows. 
Firstly, in this section, we explain the scheme only focusing on envelope-modification for simplicity. 
This is because such envelope-modification is the central topic throughout most of this paper. 
It is however possible for us to consider modifications of chemical compositions in regions 
other than the envelope; one of such applications can be found in Subsection \ref{sec:4-2-1} and Section \ref{sec:4-3}, 
where not the envelope but the deep radiative region is modified to render the chemical composition 
gradient be artificially steeper. 

Secondly, in this study, 
%chemically 
envelope-modified models are constructed by exchanging hydrogen for helium (mimicking helium enhancements in the envelope) 
so that the %mass coordinate is fixed, i.e. $\delta m = 0$. 
total mass is fixed, i.e. $\delta M = 0$. 
%Therefore, we have to be careful when we consider a progenitor of a certain envelope-modified model. %; 
Therefore, we should be aware that an unmodified model is not identical to a progenitor of 
%chemically 
envelope-modified models. %; 
%there are in principle an infinite number of possible progenitors with different masses 
%from which the envelope-modified model can be computed 
%by ``correctly'' modifying the progenitors with the corresponding $\delta m$. 
On the other hand, the fixed total mass enables us to simplify the non-standard modeling of the star, 
and we can carry out g-mode and p-mode fittings independently 
from each other 
%a scheme of non-standard modeling of the star 
as will be shown in Section \ref{sec:3}. 
We discuss the point later again in Section \ref{sec:4-1} as well. 
%%Important point is that what we would like to know is not the exact amount of mass accreted star but whether the star has experienced such envelope modification during the evolution or not!!
%\subsection{Envelope-modifying scheme} \label{sec:2-1}

\subsection{Concept for envelope-modifying scheme} \label{sec:2-1}
\begin{figure} [t]
 \begin{center}
  \includegraphics[scale=0.35]{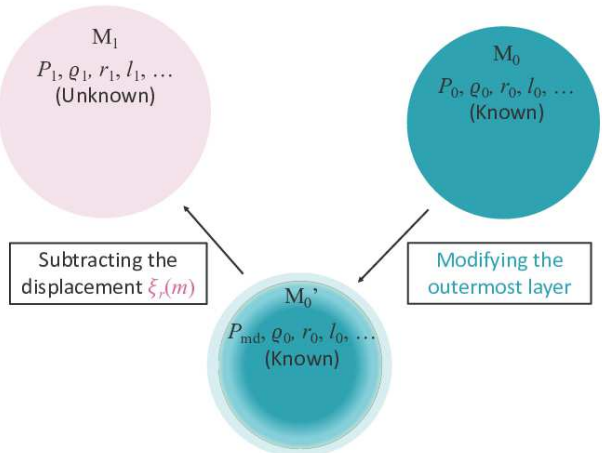}
  \caption{\footnotesize Schematic view of the three models and the relations among them.}
  \label{three_models}
 \end{center} 
\end{figure}

Generally speaking, a model of a star at a certain age is considered as 
a sphere whose interior is divided into multiple spherical layers called mass shells. 
The mass inside the concentric sphere $m$ 
%(see the definition in the Appendix A: UNDER CONSTRUCTION) 
is often taken as an independent variable. 
Dependent variables as functions of $m$ (strictly speaking, they are also functions of time which is assumed to be fixed here), 
namely, the distance from the center of the star $r$, the pressure $P$, the temperature $T$, 
the local luminosity $l$, and the mass fractions for each chemical element $X_{i}$ 
(the index $i$ represents each chemical element), 
together with other parameters such as the thermodynamic quantities 
(the density $\rho$, the adiabatic sound speed $c^2$, etc.), 
the opacity $\kappa$, and the nuclear energy generation rate $\varepsilon$ are 
assigned for each mass shell so that the set of the equations 
for stellar structure and evolution \citep[see, in particular, Section 6 in][]{Paxton2011} %(see the Appendix A: UNDER CONSTRUCTION) 
is satisfied with the variables above with certain precisions. 
This is the standard view of a stellar model, and let us call this the model $M_{0}$ (see Figure \ref{three_models}).

Then, how do we model the processes of the chemical composition modification, 
which is possibly caused by, for instance, mass accretion, 
and incorporate the effects on the structural variables $(r, P, T, l, X_{i})$ 
of the stellar model $M_{0}$ defined in the last paragraph? 
We simplify the processes with four steps as described in the following paragraphs. 

First, we determine a particular mass shell in the model above which the chemical compositions $X_{i}$ are to be changed. 
The amount of the modification to $X_{i}$ is arbitrary, 
and an example of the explicit forms for the amount is given in Section \ref{sec:2-3} as a function of $m$. 
We fix the structural parameters other than $X_{i}$ (namely, $r$, $P$, $T$, and $l$). The thermodynamic parameters, the opacity, 
and the nuclear energy generation rate %, and so on 
are accordingly changed based on the already computed tables such as OPAL. 
Note that there are usually two degrees of freedom in terms of the variation in thermodynamic quantities when $X_{i}$ are changed. 
In this study, the density $\rho$ (in other words, the distance of the mass shell from the center $r$) 
and the temperature $T$ are assumed to be fixed. 
We can instead select the other sets of parameters to be fixed, for example, 
the specific entropy $s$ and the density $\rho$, which nevertheless do not strongly affect the final results in this study. 

The modification introduced in the first step results in the change of the mean molecular weights $\mu$ of the modified mass shells, 
also leading to the change of the pressure $P$. This spherical model $M_{0}$ is thus no longer in hydrostatic equilibrium state; 
the hydrostatic equation is not satisfied with the new set of modified parameters (remember that $r$ is assumed to be fixed, and thus the 
gravitational force is not changed though the pressure gradient is). 
Let us call this model the perturbed model $M_{0}'$ (see Figure \ref{three_models}). 

Then, the perturbed model $M_{0}'$ should start radially oscillating motions around 
%\edit1{evolving toward} %radially oscillating around 
a particular hydrostatic equilibrium point. 
Based on the assumption of the adiabatic process for the oscillating motions 
(which can be partly justified because the 
dynamical timescale is usually much smaller 
than the thermal relaxation timescale for a main-sequence star), 
there is one unique equilibrium point, and 
we adopt the point as the envelope-modified model (denoted as $M_{1}$ in Figure \ref{three_models}). 
Specifically, the model $M_{1}$ can be obtained by 1) solving the second order differential equation %formulated 
%based on 
which is similar to that for the linear adiabatic radial oscillation 
%\edit1{except that the acceleration term is replaced with a term related to 
except that the acceleration term is replaced with %now arises from 
the deviation from the hydrostatic equilibrium (see Section \ref{sec:2-2}) %\edit1{the theory of linear adiabatic perturbations} %
and 2) subtracting the thus determined radial displacements from the radial coordinates of the perturbed model $M_{0}'$. 
%%%\edit1{, and %\edit1{with the dynamical timescale} 
%(denoted as $M_{1}$ in Figure \ref{three_models}) 
%%%we adopt the new equilibrium point as the envelope-modified model (denoted as $M_{1}$ in Figure \ref{three_models}). }
%%Although the radially oscillating motions around $M_{1}$ is different 
%%from any radial eigenoscillations of $M_{0}$ (since the chemical compositions of the models are different), 
%(if this is the case, $M_{1}$ and $M_{0}$ should be identical to each other), NO!!
%and we will use the term ``oscillation'' with the 
%former meaning in the rest parts of this section. 
%(though we will later see that the model $M_{1}$ can be determined 
%in a manner similar to that for the linear adiabatic radial oscillation. 
%We also would like to emphasize that what we focus on is 
%
%to which we attempt to resettle the model $M_{0}'$. 
%This procedure can be achieved by considering that the perturbed state is caused by 
%radial displacements added to another hydrostatic equilibrium state 
%different from the unperturbed model. 
%i}
%%%If we assume the model $M_{0}'$ to be adiabatically \edit1{perturbed from the model $M_{1}$} %\edit1{pulsating} %\edit1{evolving} %oscillating 
%%%(which can be partly justified because the 
%%%dynamical timescale is usually much smaller than the thermal relaxation timescale for a main-sequence star), 
%which is explained in Section \ref{sec:2-2}. 
%In other words, we here consider 
This is the second step.

In the second step, the modification is implicitly 
assumed to be small enough that we can treat the modification as a perturbation, which is 
required to guarantee the validity of the radial displacements determined by solving the linear differential equation. 
Therefore, to obtain a model whose outer region is significantly modified 
compared with the unperturbed model $M_{0}$, 
we have to repeat the step 1 and 2 substantial times. 
We denote the model obtained in this way as $M_{1}'$. 

Finally, because the model $M_{1}'$ is considered to have deviated from a thermal equilibrium state, we have to 
resettle the model again toward a thermal equilibrium state. 
This final step can be done by, in this study, further evolving the model $M_{1}'$ for the 
corresponding thermal relaxation timescale of $\sim$ a few million years 
%applying the Newton-Raphson method to the model $M_{1}$' 
so that the resettled model satisfies the equation of the energy conservation and the equation for the temperature gradient. 
%In principle, this stage is somehow similar to the contraction phase of the pre-main-sequence stars for which rates of gravitational energy generation $\varepsilon_{g}$ is negative \citep[see, e.g.,][]{Kippenhahn_text}. 
%Some explanations for $\varepsilon_{g}$ can be found in Subsection \ref{3-3-2}. 

\subsection{Formulation} \label{sec:2-2} %based on the linear adiabatic radial oscillation}
%\edit1{\subsection{Formulation based on the theory of linear adiabatic perturbations} \label{sec:2-2}}
%for simplicity, mass coordinate is fixed. but the formulation, not so changed. 
The mathematical formulations to describe the steps in the previous section are presented in this section. 
Let us start with the set of equations for stellar structure 
which the parameters of the unperturbed model $M_{0}$ satisfy, expressed as below: 
%and denote the stellar parameters as $q_{0}$ (here all the parameters are represented by $q$.) Because these parameters have to satisfy equations (\ref{Eq001}) to (\ref{Eq004}), we have the following relations
\begin{equation}
\frac{dP_{0}}{dm} = -\frac{Gm}{4 \pi r_{0}^4}, \label{Eq_M0_hydro} 
\end{equation} 
\begin{equation}
\frac{dr_{0}}{dm} = \frac{1}{4 \pi r_{0}^2 \rho_{0}}, \label{Eq_M0_mass} 
\end{equation} 
\begin{equation}
\frac{dT_{0}}{dm} = -\frac{Gm}{4 \pi r_{0}^4}\frac{T_{0}}{P_{0}} \nabla_{0}, \label{Eq_M0_temp} 
\end{equation} 
and
\begin{equation}
\frac{dl_{0}}{dm} = \varepsilon_{n,0} - \varepsilon_{\nu,0}, \label{Eq_M0_L} 
\end{equation} 
where the subscripts $0$ are representing the unperturbed state of the model $M_{0}$. 
The actual temperature gradient $\nabla_{0}$ is defined as $d \, \mathrm{ln}\, T_{0}$/$d \, \mathrm{ln}\, P_{0}$. 
The other parameters have the same meaning as in the last section. 

In the first step, the outer envelope is modified, % to some extent, 
i.e. we artificially add a small perturbation to the chemical composition $\mu$ of the unperturbed model $M_{0}$ 
\begin{equation}
\mu_{0} \to \mu_{1} = \mu_{0} + \delta \mu. \nonumber  \label{Eq_ptb_mu} 
\end{equation} 
It is totally up to us to decide how to modify the envelope. 
In this study, as shown in Section \ref{sec:2-3}, we exchange hydrogen with helium
so that %the mass coordinate $m$ is unchanged. 
the total mass $M$ is unchanged. 
The corresponding explicit form for $\delta \mu$ is given there. 

We also assume that the temperature and the density are the same as those of the starting model $M_{0}$. 
Based on these assumptions, we can calculate the perturbed pressure $P_{\rm{md}}$ by interpolating tables of equation of state such as OPAL, 
\begin{equation}
P_{0} = P(\rho_{0}, T_{0}, \mu_{0}) \to P_{\mathrm{md}} = P(\rho_{0}, T_{0}, \mu_{1}). \nonumber \label{Eq_ptb_p} 
\end{equation} 
It is then obvious that the perturbed model $M_{0}'$ is not in a hydrostatic equilibrium state as described in Section \ref{sec:2-1}. 

In the second step, we consider that the deviation from the hydrostatic equilibrium state is caused by adding the radial displacement $\xi_{r}$ 
to another hydrostatic equilibrium model $M_{1}$ (see Figure \ref{three_models}). 
The structural parameters of the model $M_{1}$ must satisfy the hydrostatic equation: 
\begin{equation}
\frac{dP_{1}}{dm} = -\frac{Gm}{4 \pi r_{1}^4}, \label{Eq_M1p_hydro} 
\end{equation} 
where the subscripts $1$ are representing the model $M_{1}$. 
%and
%\begin{equation}
%\frac{dT_{1}}{dm} = -\frac{Gm}{4 \pi r_{1}^4}\frac{T_{1}}{P_{1}} \nabla_{1}. \label{Eq_M1p_mass} 
%\end{equation} 
We can relate the new parameters to those of $M_{0}'$ as
\begin{equation}
r_{1} = r_{0} - \xi_{r}, \label{Eq_ptb1} 
\end{equation} 
\begin{equation}
P_{1} = P_{\rm{md}} - \delta P, \label{Eq_ptb2} 
\end{equation} 
\begin{equation}
T_{1} = T_{0} - \delta T, \label{Eq_ptb3} 
\end{equation} 
and 
\begin{equation}
\rho_{1} = \rho_{0} - \delta \rho. \label{Eq_ptb4} 
\end{equation} 
If we substitute relations (\ref{Eq_ptb1}) to (\ref{Eq_ptb4}) for expression (\ref{Eq_M1p_hydro}), % and (\ref{Eq011}), 
we have the following equation %s
\begin{equation}
\frac{d(P_{\rm{md}} - \delta P)}{dm} = -\frac{Gm}{4 \pi (r_{0}-\xi_{r})^4}. \label{Eq_M1_hydro} 
\end{equation} 
%and
%\begin{equation}
%\frac{d(T_{0} - \delta T)}{dm} = -\frac{Gm}{4 \pi (r_{0}-\xi_{r})^4}\frac{T_{0}-\delta T}{P_{\rm{md}}-\delta P} (\nabla_{\rm{md}} -\delta \nabla). \label{Eq017} 
%\end{equation} 
Note that we are not considering perturbed equations for the temperature gradient, 
because we concentrate on the adiabatic process, as it is explained later. 
%it is relatively simple to  
%For an alternative scheme where 
%the equation of the temperature gradient is taken into account as well can be found in Appendix \ref{AppB}. 
% (UNDER CONSTRUCTION). 
The equation of the energy conservation is also not considered here 
due to the assumption that the local luminosity $l$ is fixed. See Section \ref{sec:2-1} for how to 
handle possible deviations from the thermal equilibrium states. 
%The assumption possibly results in a deviation from the thermal equilibrium state, which 
%is to be  %, and that $l$ is strongly dependent on 
%the innermost region to which the modification is added in this case; $\delta l =0$. 
%temperature gradient is modified in accordance with the change of the pressure, and it is thus denoted by $\nabla_{\rm{md}}$. 

Assuming that the perturbations are small enough 
to justify neglecting the perturbed quantities of higher than the first order, the equation above can be further simplified as below: 
\begin{equation}
 - \frac{d (\delta P)}{dm} = -\frac{Gm}{4 \pi r_{0}^4} \frac{4 \xi_{r}}{r_{0}} -\delta h,  \label{Eq_M1p_ptb_hydro} 
\end{equation} 
%and
%\begin{equation}
%-\frac{d( \delta T)}{dm} = -\frac{Gm}{4 \pi r_{0}^4}\frac{T_{0}}{P_{\rm{md}}} \nabla_{\rm{md}} \biggl (  \frac{4\xi_r}{r_{0}} -\frac{\delta T}{T_{0}} + \frac{\delta P}{P_{\rm{md}}} -\frac{\delta \nabla}{\nabla_{\rm{md}}} \biggr ) -\delta th, \label{Eq017-1} 
%\end{equation} 
where $\delta h$ is defined as %and $\delta th$ are defined as 
\begin{equation}
\delta h \equiv \frac{dP_{\rm{md}}}{dm} + \frac{Gm}{4 \pi r_{0}^4}, \label{Eq_del_h} 
\end{equation} 
%and
%\begin{equation}
%\delta th \equiv \frac{dT_{0}}{dm} + \frac{Gm}{4 \pi r_{0}^4} \frac{T_{0}}{P_{\rm{md}}} \nabla_{\rm{md}}  \label{Eq017-2} 
%\end{equation} 
which represents a degree of deviation from a hydrostatic equilibrium. 

The perturbed (thermodynamic) quantities are dependent on how we take the pathway from $M_{1}$ to $M_{0}'$. 
We discuss the simplest way where the adiabatic process is assumed. 
%For another way to deal with it, see Appendix \ref{AppB}. % B (UNDER CONSTRUCTION). 
% (I think it is better to mention that the luminosity is fixed throughout this report; we assume that the core region is well-constrained from the observed mean g-mode period spacing.)
In the adiabatic process, there is no heat transfer among the mass shells of the model. 
We can relate the small perturbations of the thermodynamic quantities 
such as $\delta P$, $\delta T$, and $\delta \rho$ to their values $P_{\rm{md}}$, $T_{0}$, and $\rho_{0}$ with adiabatic exponents as follows:
%
%\subsection{Solving $\xi_{r}$ and $\delta T$ assuming adiabatic process}
%\label{1-2-1}
%The simplest case is the adiabatic process in which there is no heat transfer among the mass shells of the model. We can relate the small perturbations of the thermodynamic quantities such as $\delta P$, $\delta T$, and $\delta \rho$ to their values $P_{\rm{md}}$, $T_{0}$, and $\rho_{0}$ with adiabatic exponents as follows:
\begin{equation}
\frac{\delta P}{P_{\rm{md}}} = \Gamma_{1} \frac{\delta \rho}{\rho_{0}} \label{Eq_ptb_gamma1} 
\end{equation} 
and 
\begin{equation}
\frac{\delta T}{T_{0}} = (\Gamma_{3}-1) \frac{\delta \rho}{\rho_{0}}, \label{Eq_ptb_gamma3} 
\end{equation} 
where the two adiabatic exponents are defined as 
\begin{equation}
 \Gamma_{1} \equiv \biggl ( \frac{\partial \  \rm{ln} \  \it{P}}{ \partial \  \rm{ln} \ \rho}  \biggr )_{\rm{ad}} \nonumber \label{Eq_gamma1} 
\end{equation} 
and
\begin{equation}
 \Gamma_{3}-1 \equiv \biggl ( \frac{\partial \  \rm{ln} \  \it{T}}{ \partial \  \rm{ln} \ \rho}  \biggr )_{\rm{ad}}, \nonumber \label{Eq_gamma3} 
\end{equation} 
and they can be obtained from tables for equation of state. 

When we insert expression (\ref{Eq_ptb_gamma1}) into equation (\ref{Eq_M1p_ptb_hydro}), it can be rewritten as 
\begin{equation}
-\frac{d}{dm}(c^2_{\rm{md}} \delta \rho) = -\frac{Gm}{4 \pi r_{0}^4} \frac{4 \xi_{r}}{r_{0}} -\delta h. \label{Eq_osc_c} 
\end{equation} 
The adiabatic sound speed for the model $M_{0}'$ is expressed as $c^2_{\rm{md}}$. 
Let us then consider the mass conservation in the case of the linear oscillation, namely, %\edit1{perturbations}, namely, 
\begin{equation}
\rho' + \nabla \cdot (\rho_{0} \boldsymbol{\xi}) = 0, \nonumber \label{Eq_ptb_mass} 
\end{equation} 
based on which we can relate the density perturbation $\delta \rho$ to the radial displacement $\xi_{r}$ in the following way
\begin{equation}
\delta \rho = -\rho_{0} \frac{1}{r_{0}^2} \frac{d}{d r_{0}} (r_{0}^2 \xi_{r}).  \label{Eq_delrho_xir} 
\end{equation} 
We have used a relation between the Eulerian perturbation ($\rho'$) and the Lagrangian perturbation ($\delta \rho$), 
and we adopt the spherical coordinate to articulate the specific form of the differentiation. 
For the convenience in later discussions, we express the differentiation in expression (\ref{Eq_delrho_xir}) 
in terms of the mass coordinate using the expression (\ref{Eq_M0_mass})
\begin{equation}
\delta \rho = -\frac{2\rho_{0}}{r_{0}}\xi_{r} - 4\pi r_{0}^2 \rho_{0}^2 \frac{d\xi_{r}}{dm}.  \label{Eq_delta_rho} 
\end{equation} 

Combining the expressions (\ref{Eq_osc_c}) and (\ref{Eq_delta_rho}), we finally have a linear differential equation for the radial displacement $\xi_{r}$ in the case of the adiabatic process as follows:
\begin{eqnarray}
4\pi r_{0}^2 \rho_{0}^2 c_{\rm{md}}^2 \frac{d^2 \xi_{r}}{dm^2}
 + \biggl [ \frac{d}{dm} (4\pi r_{0}^2 \rho_{0}^2 c_{\rm{md}}^2) + \frac{2\rho_{0} c_{\rm{md}}^2}{r_{0}} \biggr ] \frac{d \xi_{r}}{dm} \nonumber \\
 +   \biggl [  \frac{d}{dm} \biggl (  \frac{2\rho_{0} c_{\rm{md}}^2}{r_{0}} \biggr ) + \frac{Gm}{\pi r_{0}^5}  \biggr ] \xi_{r} + \delta h =0, \nonumber  \\\label{Eq_rad_osc} 
\end{eqnarray} 
with the following boundary conditions 
\begin{equation}
\xi_{r} = 0 \,\,\, (\mathrm{at \, the \, center}) \label{Eq_bc1} 
\end{equation} 
\begin{equation}
%\frac{d\xi_{r}}{dm}  = 0  \,\,\, (\mathrm{at \, the \, surface}). \label{Eq_bc2} 
%\delta \rho  = 0  \,\,\, (\mathrm{at \, the \, surface}), \label{Eq_bc2} 
\delta P  = 0  \,\,\, (\mathrm{at \, the \, surface}), \label{Eq_bc2} 
\end{equation} 
the latter of which is the so-called zero boundary condition. 
%See equation (\ref{Eq_delta_rho}) for the explicit form of $\delta \rho$. 

Equation (\ref{Eq_rad_osc}) can be numerically solved under the boundary conditions (\ref{Eq_bc1}) and (\ref{Eq_bc2}) 
when we have all the properties of the perturbed model $M_{0}'$. 
We can compute the density perturbation $\delta \rho$ based on expression (\ref{Eq_delta_rho}), 
and subsequently, the temperature perturbation $\delta T$ based on expression (\ref{Eq_ptb_gamma3}). 
Because the differential equation has been the linearization of the perturbed equation (\ref{Eq_M1_hydro}), 
we have to iterate the procedure explained above. 

%Note that we do not have to solve the second equation (\ref{Eq017-1}). (I have not found exact reasons why we only have to solve the first equation... Do the second equation and the first one have the same meaning under the adiabatic process (or the isothermal process)?? Or we have to solve the second equation as well and we also have to determine the perturbed values in the least-squares sense??? How about the relation between 1.2.1 and 1.2.3??) 
%
%copy!
%
After we resettle the perturbed model with one perturbation in the mean molecular weight, 
we just repeat the same procedure until we obtain the model whose envelope is as modified as we would like to 
as is shown later in Section \ref{sec:2-3}. %\ref{sec:3-3}.} 
There are thus no special mathematical formulations in the third step and the fourth step. 

\subsection{Envelope-modified models and the frequencies} \label{sec:2-3}
In this section, we present examples of envelope-modified models, which are 
characterized by helium enhancements in the envelopes, 
computed based on the scheme explained in the preceding section. 
We also show 
%p-mode 
eigenfrequencies of the envelope-modified models and a 
relation between extents of the helium enhancement and corresponding %p-mode 
frequency variations, 
which are to be utilized in the non-standard modeling of the star as will be seen in Section \ref{sec:3}. 
%This section is partly independent from the preceding ones, and there is no additional prescriptions for the scheme to be shown. 
%Instead, eigenfrequencies of an envelope-modified model which is in the hydrostatic and thermal equilibrium states are presented 
%because to compute the frequencies is one of the most prominent ingredients for carrying out asterosesimic modeling as is described in Chapter \ref{4}. 

The basic settings for the calculation follow. 
First, we prepare an unperturbed model. 
The mass and the initial helium abundance are $1.30M_{\odot}$ and $0.260$. 
The metallicity is $0.003$, and it is unaltered during the envelope-modification. 
The age of the model is determined based on the asymptotic value of the g-mode period spacing $\overline{\Delta P_{\mathrm{g}}}$
which has been frequently used as an indicator for stellar evolutionary stages \citep[see, e.g.,][]{Unno_text, Aerts_text}; 
the evolutionary calculation is stopped when 
$\overline{\Delta P_{\mathrm{g}}}$ 
(one of the outputs of MESA, computed based on the integration of the $\rm{Brunt}$-$\rm{{V\ddot{a}is\ddot{a}l\ddot{a}}}$ 
frequency of the model) is $2100 \, \mathrm{s}$, 
which is close to the mean value of the observed g-mode period spacing for KIC 11145123, 
$2070 \, \mathrm{s}$ \citep{Kurtz2014}. 
The extent of overshooting $f_{\mathrm{ovs}}$ \citep[see the definition, for example, in Section 5 of][]{Paxton2011} is set to be 0.014, 
which is around the typical values for $f_{\mathrm{ovs}}$. 
The diffusion process is activated. The default settings in MESA are used for the other prescriptions. 
The parameters determined in the above are mostly 
based on the previous models of KIC 11145123 \citep{Kurtz2014, Takada_Hidai2017}. 

\begin{figure} [t]
 \begin{center}
  \includegraphics[scale=0.32,angle=270]{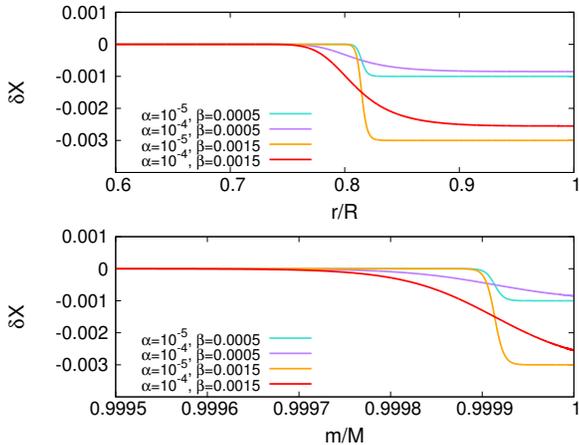}
  \caption{\footnotesize Examples of the modifications in the hydrogen mass fraction 
  computed following the equation (\ref{Eq_delta_mu}), with 
  different abscissas, namely, the fractional radius $r/R$ (top) and the fractional mass $m/M$ (bottom). %, and the mesh number (bottom). 
  %Note that the fractional mass is represented by $1-m/M$. %, which aims at rendering the readers to und
  Four ways of modification are shown changing the parameters $\alpha$ and $\beta$. 
  The parameter $m_{\mathrm{c}}$ is fixed to be $0.99976$. 
  It is seen that as $\beta$ increases the amount of the modification increases as well. 
  Increasing $\alpha$ leads to the wider %shallower 
  transition from the unperturbed region to the perturbed one. 
  }
  \label{del_mu_s}
 \end{center} 
\end{figure}
Then, the envelope-modifying scheme is applied to the model. 
The differential equation (\ref{Eq_rad_osc}) is solved under the zero boundary conditions (\ref{Eq_bc1}) and (\ref{Eq_bc2}) 
based on the second-order implicit scheme 
where the staggered mesh is adopted to describe the 
quantities and their derivatives \citep[see Figure 9 in][]{Paxton2011}. 
We parameterize the modification (hydrogen depletion or helium enhancement) in the following way: 
\begin{equation}
\delta X \equiv X_{\mathrm{md}} - X_{0} = - \beta \times  \biggl ( \mathrm{tanh} \biggl (  \frac{m-m_{\mathrm{c}}}{\alpha}  \biggr ) + 1 \biggr ), \label{Eq_delta_mu} 
\end{equation} 
in which $X_{\mathrm{md}}$ and $X_{0}$ are hydrogen mass fractions 
for the perturbed model $M_{0}'$ and the unperturbed model $M_{0}$ (see Figure \ref{three_models}). 
There are three free parameters in the expression (\ref{Eq_delta_mu}), namely, $\beta$, $\alpha$, and $m_{\mathrm{c}}$ which 
determine the extent, the width of a transition, and the depth of the modification, respectively (see Figure \ref{del_mu_s}). 
Note that the modification in the helium mass fractions $\delta Y$ is 
given as $-\delta X$ so that the sum of the mass fractions remains to be unity. 
It should be also noted that what we practically perturb is not the mean molecular weights $\mu$ but 
the hydrogen mass content $X$. 
% since $\mu$ is one of the outputs in the case of the equation-of-state module built in MESA. 

In this section, we express one modification with the following parameters: 
$\alpha=7.5 \times 10^{-4}$, $\beta=5\times10^{-6}$, and $m_{\mathrm{c}}=0.99976$. 
The modification is added to the unperturbed model 
$4\times 10^{4}$ times, and models perturbed $10^{4}$, 
$2\times 10^{4}$, $3\times 10^{4}$, and $4\times 10^{4}$ times are 
preserved for calculations of the eigenfrequencies. 
It has been confirmed that $\delta h$ of the envelope-modified models is 
at most of the order of $10^{-6}$ %(as expected from a single-precision computation) 
which is almost the same as those for ordinary stellar models computed via MESA, 
validating that our scheme is correctly working. 
The calculation of eigenfrequencies is via GYRE \citep{Townsend2013}. %, which is a linear adiabatic oscillation code. 

\begin{figure} [t]
 \begin{center}
  \includegraphics[scale=0.32,angle=270]{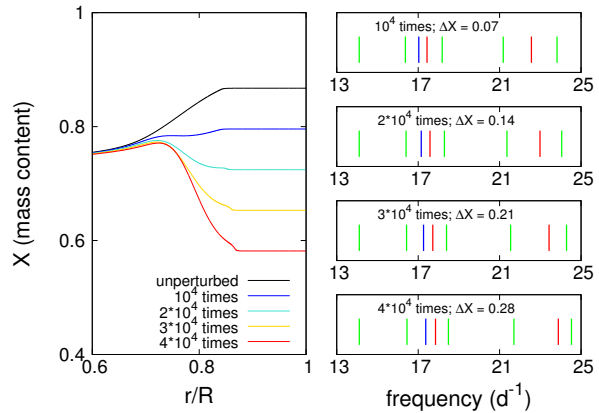}
  \caption{\footnotesize Modified hydrogen fractions (left) and the corresponding p-mode eigenfrequencies (right). 
  The envelope-modified models (left) are obtained after $10^{4}$ (blue), 
  $2 \times 10^{4}$ (turquoise), $3 \times 10^{4}$ (gold), and $4\times10^{4}$ (red) times perturbations. 
  The unperturbed profile is also indicated (black). 
  The hydrogen abundance decreases more as the envelope is perturbed more, 
  where $10^{4}$ times perturbations approximately correspond to the decrease in the surface hydrogen abundance $\Delta X\sim0.07$. 
  P-mode eigenfrequencies for radial modes (blue), dipole modes (red), and quadrupole modes (green) are shown
  for each perturbed model in the right panels. 
   %because the modification of the envelope should mainly affect not the g-mode cavity but the p-mode cavity, 
   %leading to the frequency variations for p modes. 
   It is seen that the more the envelope is modified, the more the p-mode frequencies are shifted, as expected. 
   The frequency range is chosen based on the observed p-mode frequencies of KIC 11145123 
   \citep[see also Table 4 of][]{Kurtz2014}.
  }
  \label{ptb_and_frq_var}
 \end{center} 
\end{figure}
\begin{figure} [t]
 \begin{center}
  \includegraphics[scale=0.34]{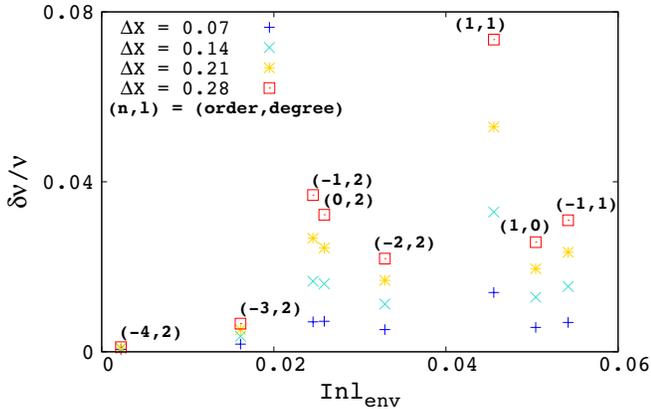}
  \caption{\footnotesize Relative frequency variation, defined as $\delta \, \mathrm{ln} \, \nu _{ij}= (\nu_{ij}-\nu_{0j})/\nu_{0j}$, 
  against the corresponding envelope mode inertia, defined in the expression (\ref{Eq_env_mode_inertia}), 
  computed based on the envelope-modified models which are here represented by 
  the surface hydrogen abundance difference $\Delta X \sim0.07$ (blue), 
  $0.14$ (turquoise), $0.21$ (gold), and $0.28$ (red). 
  (see also Figure \ref{ptb_and_frq_var}). 
  The radial order and spherical degree of each mode are presented as well. 
  %as $I_{j}^{\mathrm{env}}=\int_{\mathrm{env}} \rho |\boldsymbol{\xi}_{j}|^2 dV$. 
  Detailed meanings of the indices can be found in the text.   
  Rough proportionality of the relative frequency variation 
  with respect to the extent of modification can be confirmed. 
  A positive correlation, though non-monotonic, between the relative frequency variation and the envelope mode inertia 
  is also seen, which is discussed in the text. 
  }
  \label{dlnf_vs_inl_env_Ch3}
 \end{center} 
\end{figure}
%\begin{figure} [t]
% \begin{center}
  %\includegraphics[scale=0.32,angle=270]{dlnf_vs_inl_env_Ch3}
  %\caption{\footnotesize Relative frequency variation, defined as $\delta \, \mathrm{ln} \, \nu _{ij}= (\nu_{ij}-\nu_{0j})/\nu_{0j}$, 
  %against the corresponding envelope mode inertia, defined in the expression (\ref{Eq_env_mode_inertia}). 
  %%%%as $I_{j}^{\mathrm{env}}=\int_{\mathrm{env}} \rho |\boldsymbol{\xi}_{j}|^2 dV$. 
  %See detailed meanings of the indices in the text. 
  %Rough proportionality of the relative frequency variation 
  %with respect to the extent of modification can be confirmed. 
  %A positive correlation between the relative frequency variation and the envelope mode inertia is also seen. 
  %}
  %\label{dlnf_vs_inl_env_Ch3}
 %\end{center} 
%\end{figure}
Results of the computations are presented in Figure \ref{ptb_and_frq_var}. 
It is evident that the hydrogen is less abundant as the envelope is modified more (see the left panel of Figure \ref{ptb_and_frq_var}). 
Some of the computed abundance profiles exhibit the inversion in the mean molecular weight 
due to the envelope modification (see cyan, yellow, and red curves in the left panel of Figure \ref{ptb_and_frq_var}), 
but, in our settings for MESA, 
it has been confirmed that the modified profiles are stable at least for the thermal timescale. 
Moreover, implementing mixing processes which work in the presence of mean molecular weight inversions 
with the thermal timescale (such as thermohaline mixing) does not 
change results of the non-standard modeling of KIC 11145123 significantly. 
We therefore do not further discuss the point in this study. 
%We discuss the point later in Section \ref{sec:3-1}. 

%It is also evident in Figure \ref{ptb_and_frq_var} that the hydrogen is less abundant as the envelope is modified more, 
The hydrogen decrease represented by $\Delta X$ in Figure \ref{ptb_and_frq_var} 
(or, the helium enhancement) brought about by the envelope modifications 
should affect the adiabatic sound speed in the envelope. 
%
%It should be also noted that the modified abundance profiles are maintained even when the modified models are 
%evolved more for $\sim$ a few hundred million years (much longer than the thermal timescale). 
%
The p-mode frequencies, which are strongly dependent on the adiabatic sound speed \citep[see, e.g.,][]{Unno_text}, 
%for the models 
are thus varying as we perturb the envelope more (see the right panels in Figure \ref{ptb_and_frq_var}). 
Interestingly (and importantly), the amounts of the frequency variations are, roughly speaking, 
proportional to the amounts of the modification, 
though the dependence should be non-linear. 
%though the modification process should be non-linear. 
In addition, the amounts of the frequency variations are different from mode to mode, 
which is readily confirmed when we see the right panels in Figure \ref{ptb_and_frq_var} 
(compare, for instance, the blue bar and the red bars). 
We present a further discussion on that point in the following paragraphs, 
since these features help us reduce the size of the grid with which 
the non-standard modeling is carried out as we will see later in Section \ref{sec:3-2}. 

To see relations between the amounts of frequency variations and the mode properties more clearly, 
we plot, in Figure \ref{dlnf_vs_inl_env_Ch3}, the relative frequency variation, 
defined as $\delta \, \mathrm{ln} \, \nu _{ij}= (\nu_{ij}-\nu_{0j})/\nu_{0j}$, 
against the corresponding ``envelope mode inertia'' which is defined as below: 
\begin{equation}
I_{j}^{\mathrm{env}}=\int_{\mathrm{env}} \rho |\boldsymbol{\xi}_{j}|^2 dV, \label{Eq_env_mode_inertia} 
\end{equation} 
where the indices $i$ and $j$ stand for the extent to which 
a model is modified %with a certain extent 
%($2500$, $5000$, $7500$, or $10^{4}$ times) 
($\Delta X\sim0.07$, $0.14$, $0.21$, and $0.28$, in this case) 
and a particular mode, respectively. 
The unperturbed model is designated as $i=0$, 
and the eigenvector of the mode is expressed as $\boldsymbol{\xi}_{j}$. 
The envelope mode inertia is computed by integrating $\rho |\boldsymbol{\xi}_{j}|^2$ 
in the part of the envelope determined by us beforehand 
(here, the envelope is defined as a region $r/R>0.7$). 
Note that the total mode inertia (obtained by carrying out the integration throughout the model) is normalized to be $1/4\pi$. 
%(normalized by the total mode inertia which is here fixed to be $1/4\pi$) %(see the caption of the figure for the meanings of the variables) 

%We plot Figure \ref{dlnf_vs_inl_env_Ch3} because 
%the ratio between an envelope mode inertia and the total inertia contains information on 
%which part of the star the mode mainly propagates. 
Strictly speaking, we can describe the frequency variations 
based on the structure kernels of the unperturbed model and 
the structural differences caused by the modifications. 
But this is not the case here because the modifications are too large to consider them as small perturbations 
and apply the first-order perturbation theory to explain the frequency variations, 
which is the reason why we, as a rough approximation, have chosen $I_{j}^{\mathrm{env}}$ instead of 
structure kernels as an indicator for the sensitivity to the envelope modifications (Figure \ref{dlnf_vs_inl_env_Ch3}). 
It is seen that, for a given envelope mode inertia, the relative frequency variation is proportional to the extent of modification. 
We can also see that, for a particular extent of modification (see, e.g., a series of red squares in Figure \ref{dlnf_vs_inl_env_Ch3}), 
the relative frequency variation increases (though not exactly monotonically) 
as the envelope mode inertia increases, 
%is roughly proportional to the envelope mode inertia, which is 
which is understandable because the chemical composition modification in the envelope 
does not affect properties of a mode if the mode does not have sensitivity 
(which is represented by the envelope mode inertia) in the modified envelope. 

It should be instructive to mention that we can qualitatively explain the non-monotonic trend 
seen in Figure \ref{dlnf_vs_inl_env_Ch3} 
by checking the mode inertia densities $\rho |\boldsymbol{\xi}_{j}|^2 r^2$ of the unperturbed model. 
For example, the inertia density of the mode with $(n,l)=(1,1)$ is maximum around $r/R\sim0.95$ %in the outer envelope ($r/R>0.9$), 
which is included in a region most affected by the modifications, 
leading to the larger frequency variation compared with 
those of the other modes such as modes with $(n,l)=(-2,2)$ and $(-1,1)$ 
whose mode inertia densities are maximum around $r/R\sim 0.8$ 
where the structure is less perturbed than that around $r/R\sim0.95$, 
%
%which are mixed modes having sensitivity both in the deep region and in the outer envelope, 
leading to the relatively smaller frequency variations. 
We however do not attempt to explain all the non-monotonic signatures in Figure \ref{dlnf_vs_inl_env_Ch3} 
not only because, as shown in Section \ref{sec:3}, assuming the rough proportionality in the non-standard modeling 
is sufficient to obtain a reasonable envelope-modified model of KIC 11145123 
but also because taking the non-monotonic trend into account has little impact on the final inference of 
the non-standard modeling. 

Finally, we would like to mention that g-mode frequencies are almost insensitive to 
envelope modifications; the g-mode frequencies of the envelope-modified models are 
mostly the same. 
This is because g modes mainly propagate through the deep radiative region \citep[see, e.g.,][]{Unno_text} 
and their envelope mode inertias are almost zero. 
The properties of g- and p-mode frequency variations with respect to envelope modifications 
are to be utilized in Section \ref{sec:3} 
where asteroseismic non-standard modeling of KIC 11145123 is performed. 

\section{Non-standard modeling} \label{sec:3}
We have three steps in the non-standard modeling of the star as below. 
First, we compute ordinary models via MESA to fit atmospheric parameters %as well as possible 
with a relatively coarse grid (Section \ref{sec:3-1}). 
Then, a finer grid is constructed based on the results of the coarse-grid-based modeling to 
compute models reproducing g-mode frequencies (Section \ref{sec:3-2}).  
%fit g-mode frequencies, and 
Finally, we modify the envelopes of the models 
by increasing helium mass contents in the envelopes to fit p-mode frequencies (Section \ref{sec:3-3}). 
Such independent modeling can be achieved when, as we see in Section \ref{sec:2-3}, 
the structures of the deep regions are not affected much during the envelope modifications. 
For each section, we present the specific procedures, the parameter ranges, 
and the results. 
%After we present specific procedures and parameter ranges for 
%carrying out the non-standard modeling of KIC 11145123 (Section \ref{sec:3-1}), 
%the results are given in Section \ref{sec:3-2}. 

\subsection{Coarse-grid-based modeling \label{sec:3-1}}
%\edit1{The }
\subsubsection{Specific procedures} \label{sec:3-1-1}
We firstly prepare the following grids of parameters: 
mass $M$ ($1.1$--$2.1M_{\odot}$, with the step of $0.1M_{\odot}$ between $1.1$--$1.7M_{\odot}$ 
and with the step of $0.2M_{\odot}$ between $1.7$--$2.1M_{\odot}$), 
initial helium abundance $Y_{\mathrm{init}}$ ($0.25$--$0.27$, with the step of $0.01$), 
initial metallicity $Z_{\mathrm{init}}$ ($0.002$--$0.004$, with the step of 0.001), 
and the extent of overshooting $f_{\mathrm{ovs}}$ ($0.010$, $0.020$, and $0.030$). 
Most of the previous models are relatively low-mass stars with $M\sim1.4M_{\odot}$ \citep{Kurtz2014, Takada_Hidai2017}, 
which is the reason why the grids of the lower mass range is finer than that of the higher mass range. 
We assume that the star was born as an ordinary single star 
with the initial helium abundance of $\sim 0.26$, 
lower than that of the previous models ($Y_{\mathrm{init}}>0.30$), 
and the range for initial helium abundance is thus chosen. 
For the extent of overshooting, $f_{\mathrm{ovs}}\sim0.01$--$0.03$ is often recommended by the literatures \citep[e.g.][]{Paxton2011}. 
%, but here we include $f_{\mathrm{ovs}}=0.027$ 
%because such broader overshooting region is suggested 
%based on the analysis of the g-mode period spacing $\Delta P_{\mathrm{g}}$ pattern (see Section \ref{2-4}). 
Let us call the parameter range defined above ``the coarse grid". 

%\subsubsection{Finer grid} 
%\label{sec:3-1-2}
Then, we compute evolutionary tracks for all the points in the coarse grid. 
The evolution is stopped when the mean g-mode period spacing of the model $\overline{\Delta P_{\mathrm{g}}}$ 
(computed based on the integration of the $\rm{Brunt}$-$\rm{{V\ddot{a}is\ddot{a}l\ddot{a}}}$ frequency) 
reaches $2100 \, \mathrm{s}$, as was explained in Section \ref{sec:2-3}. 
%which is around the mean value of the observed g-mode period spacings $2070 \, \mathrm{s}$ \citep{Kurtz2014}. 
We separate the models thus computed into three groups 
based on their atmospheric parameters ($T_{\mathrm{eff}}$ and $\mathrm{log} \, g$), namely, 
the $1\sigma$ group whose models reproduce the observed atmospheric parameters 
\citep[$T_{\mathrm{eff}}=7590^{+80}_{-140} \, \mathrm{K}$ and $\mathrm{log} \, g = 4.22 \pm 0.13$ in cgs units,][]{Takada_Hidai2017} 
within $1\sigma$, 
the $2\sigma$ groups which is determined in 
the same way as the $1\sigma$ group except that the criterion is $2\sigma$, 
and the rest which consists of the models left. 

\subsubsection{Results} \label{sec:3-1-2}
 \begin{figure} [t]
 \begin{center}
  \includegraphics[scale=0.32,angle=270]{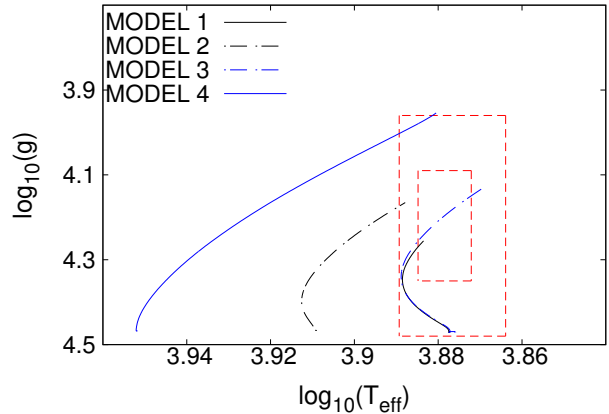}
  \caption{\footnotesize Evolutionary tracks for some of the models obtained via the coarse-grid-based modeling. 
  Models belonging to the $1\sigma$ ($2\sigma$) group are represented by black (blue) curves. 
  The set of the parameters $(M_{\odot},Y_{\mathrm{init}},Z_{\mathrm{init}},f_{\mathrm{ovs}})$ for each model 
  (in the order from ``MODEL 1'' to ``MODEL 4'') is as below: 
  $(1.2,0.25,0.002,0.020)$, $(1.3,0.27,0.003,0.010)$, 
  $(1.2,0.25,0.002,0.027)$, and 
  $(1.5,0.27,0.004,0.010)$.  
  Note that two models %(a model with $M=1.20M_{\odot}$, $Y_{\mathrm{init}}=0.25$, $Z_{\mathrm{init}}=0.002$, and $f_{\mathrm{ovs}}=0.027$  
  %and one with 
  %$M=1.20M_{\odot}$, $Y_{\mathrm{init}}=0.25$, $Z_{\mathrm{init}}=0.002$, and $f_{\mathrm{ovs}}=0.020$) 
  (MODEL 1 and MODEL 3) are 
  similar to each other; 
  the parameters are the same except for $f_{\mathrm{ovs}}$. 
  The observational uncertainties are expressed by the red dashed grids 
  following the results of \citet{Takada_Hidai2017} 
  ($T_{\mathrm{eff}}=7590^{+80}_{-140} \, \mathrm{K}$ and $\mathrm{log} \, g = 4.22 \pm 0.13$ in cgs units). }
  \label{HRD_within_2sigma}
 \end{center} 
\end{figure}
There are two main results about the coarse-grid-based modeling. 
One is that low-mass models (with masses ranging from $1.10$--$1.50M_{\odot}$) 
are favored to reproduce the observed atmospheric parameters $T_{\mathrm{eff}}$ and $\mathrm{log} \, g$; 
all of the models in either the $1\sigma$ group or $2\sigma$ group 
have masses lower than $1.5M_{\odot}$ %not depending on 
irrespective of the other parameters. 
This trend can be confirmed even when we construct the $3\sigma$ group in the same way as the other groups; 
the mass of the most massive model in the $3\sigma$ group is $1.70M_{\odot}$. 
We therefore exclude $1.7$--$2.1M_{\odot}$ from the parameter range from now on. 

Another result is that the higher value of $f_{\mathrm{ovs}}$ ($\sim0.030$) is favored 
in terms of g-mode period spacing patterns %individual g-mode frequencies 
compared with the lower ones of $f_{\mathrm{ovs}}$ ($\sim 0.010$--$0.020$). 
There is however no model in the $1\sigma$ group that favors $f_{\mathrm{ovs}}\sim0.030$. 
%Because $f_{\mathrm{ovs}}=0.027$ is suggested based on the analysis of the observed $\Delta P_{\mathrm{g}}$ pattern, 
This result implies that the models in the $1\sigma$ group are not appropriate (asteroseismically) as candidate models. 
Thus, we decided to exclude the $1\sigma$ group for further analyses. 
%(though the group is still useful for checking, for instance, the possibility that the star has evolved as a single star). 
Meanwhile, there are some models with $f_{\mathrm{ovs}}\sim0.030$ in the $2\sigma$ group, and 
we concentrate on this parameter range in the following procedures. 
Further discussions will be given later in Sections \ref{sec:3-2} and \ref{sec:4-2}. 
%The extent of overshooting for the models in the $1\sigma$ group is $f_{\mathrm{ovs}}=0.010$ or $f_{\mathrm{ovs}}=0.020$ 

Figure \ref{HRD_within_2sigma} shows some examples of evolutionary tracks of the models obtained via the coarse-grid-based modeling. 
In spite of the relatively higher $\mathrm{log} \, g\sim 4.2\pm0.1$ (cgs units), the mean of g-mode period spacings $\overline{\Delta P_{\mathrm{g}}}$ favor the TAMS stage at which stars are less dense compared with when they are on the main sequence, 
possibly leading to the preference for low-mass stellar models in the coarse-grid-based modeling. 

\subsection{Finer-grid-based modeling} \label{sec:3-2}
%\edit1{Based on the results obtained... 
%Finer grids are constructed based on the input parameters of the $1\sigma$ group 
%and the $2\sigma$ group. Let us call the newly determined parameter range ``the finer grid". }
\subsubsection{Specific procedures and results} \label{sec:3-2-1}
The results in the previous section allow us to construct a new parameter range with finer grids. 
Let us call the newly determined parameter range ``the finer grid". 
Below is the set of the finer grid: 
mass $M$ ($1.16$--$1.44M_{\odot}$, with the step of $0.02M_{\odot}$, 
initial helium abundance $Y_{\mathrm{init}}$ ($0.25$--$0.27$, with the step of $0.01$), initial metallicity $Z_{\mathrm{init}}$ ($0.002$, fixed), 
and the extent of overshooting $f_{\mathrm{ovs}}$ ($0.025$--$0.031$, with the step of $0.002$). 
The initial metallicity is fixed since all the models with $f_{\mathrm{ovs}}\sim0.030$ in the $2\sigma$ group have $Z=0.002$. 

%Finer grids are constructed based on the input parameters of the $1\sigma$ group and the $2\sigma$ group. 
%The details are shown in the next section \ref{sec:3-2} 
%(because we do not know where should be finer grids until results of the coarse-grid-based modeling are obtained). 
We again compute evolutionary tracks for the finer grid 
to obtain candidate models whose envelopes are to be modified to fit the observed p-mode frequencies. 
Evolution is stopped first when $\overline{\Delta P_{\mathrm{g}}}$ of the model reaches $2150 \, \mathrm{s}$, 
which is slightly above the mean of the observed g-mode period spacing of the star $2070 \, \mathrm{s}$, 
then the timestep for evolutionary calculation is changed 
from the default value (around $10^{7}$ years) to much smaller one around $10^5$ years. 
Evolution is restarted with the smaller timestep 
until $\overline{\Delta P_{\mathrm{g}}}$ reaches $1950 \, \mathrm{s}$, 
and all the equilibrium models computed along the evolution between $\overline{\Delta P_{\mathrm{g}}}=1950 \, \mathrm{s}$ 
and $\overline{\Delta P_{\mathrm{g}}}=2150 \, \mathrm{s}$ are saved, for 
which the corresponding eigenfrequencies (of both p and g modes) are computed via GYRE. 

\begin{figure} [t]
 \begin{center}
  \includegraphics[scale=0.32,angle=270]{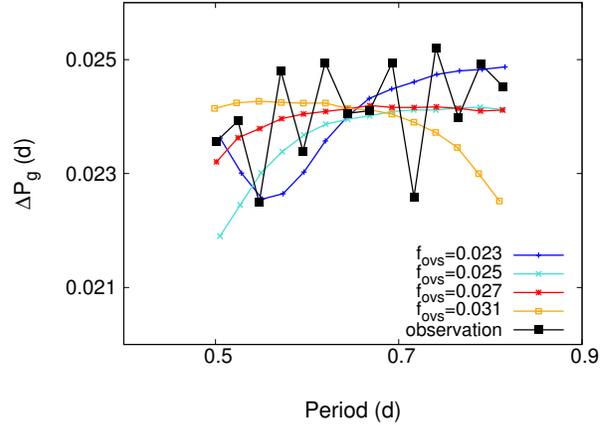}
  \caption{\footnotesize G-mode period spacing patterns of KIC 11145123 (black), 
  one of the pre-candidate models (red), and non pre-candidate models (blue, turquoise, and gold). 
  All of these models have the same mass ($1.36M_{\odot}$), initial helium abundance ($0.26$), and 
  initial metallicity ($0.002$), but with different extents of overshooting ($f_{\mathrm{ovs}}=0.023$--$0.031$) 
  which mainly results in the different behaviors of the modeled g-mode period spacing patterns. 
  It is seen that the pre-candidate model (red) can reproduce the general positive trend of the 
  observed g-mode period spacing pattern, leading to the smallest residuals (between the modeled and observed 
  g-mode frequencies) among these models.  }
  \label{dPgs_candidates}
 \end{center} 
\end{figure}

Among a series of evolutionary models for a certain set of the parameters 
($M$, $Y_{\mathrm{init}}$, $Z_{\mathrm{init}}$, and $f_{\mathrm{ovs}}$), 
the model which minimizes the sum of the squared residuals 
(normalized by the observational uncertainties) 
between the modeled and the observed g-mode frequencies is chosen 
as a candidate for ``candidate models". 
Let us call them ``pre-candidate models". 
Figure \ref{dPgs_candidates} shows modeled g-mode period spacing pattern of one of the pre-candidate models (red) 
and those of non pre-candidate models (blue, turquoise, and gold), 
compared with the observed one (black), where it is readily seen that the extent of overshooting $f_{\mathrm{ovs}}=0.027$ (red) is 
most appropriate to reproduce the gradual positive trend of the observed g-mode period spacing pattern. 
We will discuss the g-mode period spacing pattern in more detail in Section \ref{sec:4-2}. 

The modeled p-mode frequencies are subsequently checked 
to determine which pre-candidate models are appropriate for ``candidate models". 
As is described in Section \ref{sec:2-3}, the amount of the frequency variations 
caused by envelope modifications via our non-standard scheme 
is seemingly proportional to a ratio of the envelope mode inertia 
to the total inertia. 
We exploit the feature to select candidate models in the following steps. 
Though the assumption of proportionality for frequency variations is a rather crude one (see Figure \ref{dlnf_vs_inl_env_Ch3}), 
ignoring the non-monotonic trend has little impact on the final inference of the non-standard modeling 
as was described in Section \ref{sec:2-3}. 

First, a modeled radial-mode frequency which is closest to 
the observed frequency of the singlet ($\nu=17.9635133 \pm 5\times 10^{-7} $ $d^{-1}$) is chosen. 
Then, the difference between them $(\Delta \nu_{0})_{\mathrm{mod}-\mathrm{obs}}$ is computed, 
%(be careful not to be confused by the expression, this is not the large separation $\Delta \nu$, see, e.g., Subsection \ref{1-2-3}), 
which ideally becomes zero after we suitably modify the envelope of the model. 
Based on the assumption of the proportionality for the frequency variation 
caused by the envelope modification in addition to the difference $(\Delta \nu_{0})_{\mathrm{mod}-\mathrm{obs}}$, 
we can calculate an expected frequency variation for each mode as follows: 
\begin{equation}
(\Delta \nu_{i})_{\mathrm{expect}} = \frac{I_{i}^{\mathrm{env}}}{I_{0}^{\mathrm{env}}} \,(\Delta \nu_{0})_{\mathrm{mod}-\mathrm{obs}}, \label{Eq_deltap_for_cand} 
\end{equation} 
where $I_{j}^{\mathrm{env}}$ is defined as in equation (\ref{Eq_env_mode_inertia}), 
and it can be computed with the outputs of GYRE. 
For more information, see, e.g., \citet{Aerts_text}. 
Finally, we compare the expected frequency variation for a particular mode with the difference between one of the observed frequencies 
and its closest modeled frequency $(\Delta \nu_{i})_{\mathrm{mod}-\mathrm{obs}}$, 
namely, we compute the following quantities for each detected peaks: 
\begin{equation}
\biggl (  (\Delta \nu_{i})_{\mathrm{expect}} -  (\Delta \nu_{i})_{\mathrm{mod}-\mathrm{obs}} \biggr )^2,  \label{Eq_deltap_res} 
\end{equation} 
where the number of modes used in this fitting procedure is six including the singlet. 

By imposing an arbitrary criterion for the sum, 
several models which render the sum of the quantities (\ref{Eq_deltap_res}) 
below the criterion are chosen as ``candidate models". 
%The specific value of the criterion can be found in Subsection \ref{sec:3-2-2}. 
%
%Based on the grids, we carry out the finer-grid-based modeling as described in detail in Section \ref{sec:3-1}. 
We adopt a criterion, %(expressed as the sum of the quantities \ref{Eq_deltap_res}), 
above which the corresponding models are discarded and not taken as candidate models, 
so that about a tenth of pre-candidate models is chosen as a candidate model, 
which corresponds to $0.39$ in this study. 
With this criterion, we have selected five models as candidate models to which the envelope-modifying scheme is applied. 

\subsection{Envelope-modifying modeling} \label{sec:3-3}
%\edit1{Finally, ...}
%Before moving on to the envelope-modifying modeling, 
%we have a subtle step where models are computed with the same sets of the parameters 
%as those of the selected candidates except for the ages; 
%the newly computed models are younger than the original candidate models by thermal timescale for the models. 
%This is because we have to evolve envelope-modified models 
%until they settle to the thermal equilibrium states, 
%and it is possible that the evolution leads to deviations in fitted g-mode frequencies. 
%Such deviations can be avoided by using the younger candidate models. 
%
\subsubsection{Specific procedures} \label{sec:3-3-1}
The next thing we have to do is to modify the envelopes of the ``candidate'' models. 
But before moving on to the envelope-modifying modeling, 
we have a subtle step. 
%Though we do not take into account temporal evolution of the modified models 
%during the modification process, 
It should be noticed that, in our scheme, the envelope-modified models are finally evolved 
for their thermal timescales to resettle the thermal equilibrium states (see Section \ref{sec:2-1}), 
which leads to a change in structures of the deep radiative regions as well as the 
g-mode frequencies. 
Therefore, the fitted g-mode frequencies for the ``candidate'' models could 
deviate from the observed ones in the case of the envelope-modified model eventually obtained after 
evolving for the thermal timescale. 
To avoid such deviations, we compute models with the same sets of parameters 
as those of the ``candidate'' models except for the ages; 
the newly computed models are younger than the original candidate models by the thermal timescales for the models. 
Let us call them ``younger-candidate'' models. 
Note that, from now on, only the younger-candidate models are modified. 

Then, the envelopes of the chosen ``younger candidate'' models are 
gradually modified changing the parameters describing the modification, namely, 
the extent, the width of a transition, and the depth of the modification (see the expression \ref{Eq_delta_mu}). 
%As it is pointed out in Section \ref{sec:2-3}, 
For every five modifications, the eigenfrequencies of the corresponding envelope-modified model are computed via GYRE. 
It should be noted that deep inner regions are fixed and not modified so that 
the already fitted g-mode frequencies in previous steps would not be changed. 
%In other words, 
%This step has been repeated with different sets of parameters in the determined parameter range. 
Among the envelope-modified models thus calculated, ones reproducing the observed frequencies best are selected as the best models. 
%The final procedure for the envelope-modifying modeling is the same as demonstrated in Section \ref{3-5}, 
%we do not intend to repeat it here. 

%\subsection{Results} \label{sec:3-2}
% $0.39$ as the criterion (expressed as the sum of the quantities \ref{Eq_deltap_res}) 
%above which the corresponding models are discarded, and not taken as candidate models. 
%The criterion is determined so that about a tenth of pre-candidate models is chosen as a candidate model. 

%(HAVE TO MENTION CHI VALUES OF G-MODE FREQUENCIES FOR THE CANDIDATE MODELS)

\subsubsection{Results} \label{sec:3-3-2}
The envelope-modifying scheme is applied for the five candidate models, 
and we end up with a tentatively best model (within the non-standard scheme) demonstrated as below. 
The set of the parameters of this model is $M=1.36M_{\odot}$, $Y_{\mathrm{init}}=0.26$, 
$Z_{\mathrm{init}}=0.002$, $f_{\mathrm{ovs}}=0.027$, and $ \mathrm{Age}=2.169 \times 10^{9}$ years old. 
The parameters for the modifications are $r_{\mathrm{c}}=0.6$, $\alpha=5\times10^{-3}$, 
and the number of modifications is $115$ which 
corresponds to $\Delta X \sim 0.06$ ($\Delta X$ is a difference in hydrogen abundance 
between the candidate model and the modified model) at the surface. 
The logarithm of the surface gravitational acceleration and 
that of the effective temperature of the model are $3.9$ (cgs units) and $3.87$. 

\begin{figure} [t]
 \begin{center}
  \includegraphics[scale=0.32,angle=270]{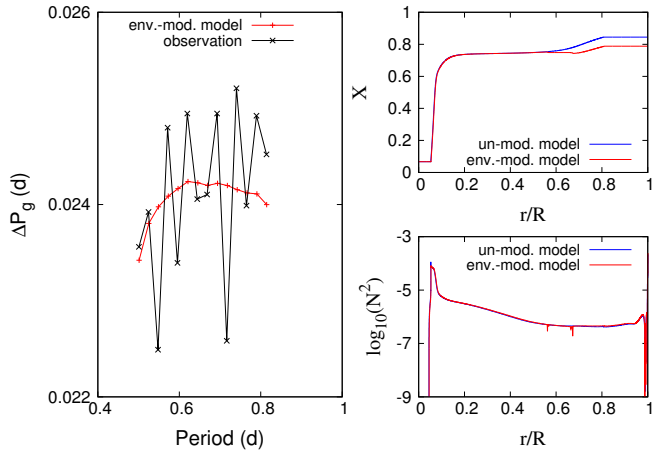}
  \caption{\footnotesize  $\Delta P_{\mathrm{g}}$ pattern for the model 
  obtained based on the non-standard modeling (red, denoted as ``env.-mod. model'') 
  and that for the observation (black). 
  The model successfully reproduces the generally positive slope 
  for periods smaller than $0.6 \,d$ of the observed $\Delta P_{\mathrm{g}}$ pattern, but 
  the other oscillatory component with a shorter period ($\Delta n \sim$ a few) cannot be reproduced with this model. 
  Right panels show the internal structures (namely, the hydrogen profile in top panel and the 
  $\rm{Brunt}$-$\rm{{V\ddot{a}is\ddot{a}l\ddot{a}}}$ frequency in bottom panel) 
  of the envelope-modified model (red) and those of the corresponding unmodified model 
  (blue, denoted as ``un-mod. model''). 
  %Right panels show the hydrogen profiles (top right) and the $\rm{Brunt}$-$\rm{{V\ddot{a}is\ddot{a}l\ddot{a}}}$ frequencies (bottom right) 
  %of the model for the whole interior of the model (top) and 
  %for an expanded look into a region just above the convective core 
  %where the chemical composition gradient develops. 
  We see two dips in the $\rm{Brunt}$-$\rm{{V\ddot{a}is\ddot{a}l\ddot{a}}}$ frequency of the envelope-modified model (bottom right), 
  which are caused by the envelope modification 
  implemented to the scheme of non-standard modeling. 
  These features nevertheless do not affect the g-mode frequencies. }
  \label{dPg_N_tentative1}
 \end{center} 
\end{figure}

The sum of the squared residuals 
(between the model and the observation) normalized by the observed uncertainties 
for g-mode frequencies is significantly smaller ($\sim 3\times 10^{5}$) 
than that in the case of the previous studies \citep[e.g.][]{Kurtz2014} ($\sim 10^{6}$). 
This improvement is brought about by the fact that we have attempted to fit the individual g-mode frequencies 
considering the extent of overshooting in contrast to the previous studies 
where only the mean value of the observed g-mode period spacing was fitted \citep[e.g.][]{Kurtz2014}. 
We can see the signature in the g-mode period spacing ($\Delta P_{\mathrm{g}}$) pattern of the envelope-modified model 
computed via GYRE, which successfully reproduces the observed positive trend 
the previous models did not (Figure \ref{dPg_N_tentative1}). % which 
%is thought to be caused by the overshooting (see Figure \ref{dPg_N_tentative1}) (also see Section \ref{2-4}). 
It is also seen that there is still a significant discrepancy between the observed $\Delta P_{\mathrm{g}}$ and the modeled one, 
especially with respect to the oscillatory component with a short period of $\Delta n\sim$ a few, where 
$n$ denotes the radial order of a certain g mode. 
We discuss the point later in Section \ref{sec:4-2}. 

\begin{figure} [t]
 \begin{center}
  \includegraphics[scale=0.33]{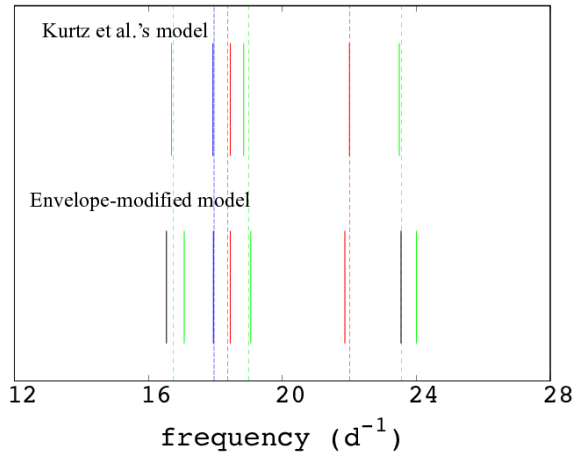}
  \caption{\footnotesize Comparison of the modeled frequencies (solid lines) with the observed ones (dashed lines) for 
  the envelope-modified model obtained based on the non-standard modeling (bottom) and for the model of \citet{Kurtz2014}. 
  The radial, dipole, quadrupole, and octupole modes are represented by blue, red, green, and black, respectively. 
  Note that the colors of the observed frequencies are based on the mode identification by \citet{Kurtz2014}. 
  In the case of the envelope-modified model, octupole modes are sometimes better 
  to reproduce the observed frequencies than the dipole-mode frequencies of the model, 
  implying the possibility of a different mode identification. 
  See the text for more discussions. }
  \label{comp_Kurtz_tent_mod}
 \end{center} 
\end{figure}
Figure \ref{comp_Kurtz_tent_mod} shows the comparison of the modeled p-mode frequencies with the observed ones. 
The radial- (blue), dipole- (red), and quadrupole- (green) mode frequencies are presented. 
For our envelope-modified model, $l=3$ modes (black) are also illustrated. 
The envelope-modified model obtained in this study fits the observed radial-mode frequency better than the model of \citet{Kurtz2014}. 
The other p-mode frequencies, however, are not fitted so well in the case of the envelope-modified model, 
especially when we follow the mode identification 
adopted in \citet{Kurtz2014} (see the caption of Figure \ref{comp_Kurtz_tent_mod} for more details). 
The mean deviation, which is defined as the mean of the absolute value of the difference 
between the modeled frequencies and the modeled ones, 
for our model is $0.2$ $d^{-1}$ while that of the best model in \citet{Kurtz2014} is $0.1$ $d^{-1}$. 

However, we can at least claim that we have partly succeeded in constructing a comparable model with previous models; 
mean deviations of some of the previous models are sometimes larger than $0.5$ $d^{-1}$ \citep[e.g.][]{Takada_Hidai2017}. 
We also would like to emphasize that the starting points are totally different 
for the current envelope-modified model and the previous models; 
in previous studies, we need to adopt too-high initial helium abundance of 
$\sim 0.3$--$0.4$ to explain the observed p-mode frequencies, but in this study, 
we can explain the observed ones with a much more ordinary initial helium abundance of 
$\sim 0.26$ by modeling the star in the non-standard manner where helium mass contents are suitably increased 
in the envelope. 

Interestingly, when we include $l=3$ modes in the mode identification, 
the mean deviation for our model reduces to be $0.1$ $d^{-1}$ 
which is comparable to that of \citet{Kurtz2014}. 
Although the mode identification of \citet{Kurtz2014} is reasonable in a sense that 
the spherical degree assigned to a certain mode is compatible with the number of the observed multiplets for the mode 
(if this is not the case, we need additional mechanisms to explain the reason why some modes 
in the multiplet are excited and the others are not), 
obtaining a hint for another possible way of mode identification for the star including $l=3$ modes could lead to 
further investigations that would, for example, challenge 
the theory of the mode excitation mechanism in $\delta$ Scuti stars, 
which has been definitely one of the unsettled subjects in asteroseismology. 

%It is, nonetheless, beyond our scope to further discuss the issue with respect to the mode identification, 
%which is a difficult subject to deal with unless we have additional observational constraints, and thus, in this study, 
%we follow the mode identification determined by \citet{Kurtz2014}. 
%It should be instructive to mention that the mode identification of \citet{Kurtz2014} is reasonable in a sense that 
%the spherical degree assigned to a certain mode is compatible with the number of the observed multiplets for the mode; 
%they assigned $l=0$ ($l=1$ or $l=2$) for the observed singlet (triplets or quintuplets). 
%This seems to be conservative compared with other cases where, for example, 
%$l=1$ is assigned for a quintuplet; 
%we need additional mechanisms to explain the reason why some modes 
%in the multiplet are excited and the others are not. 
%We discuss the point later in Section \ref{4-4}. 

\section{Discussions} \label{sec:4}
%In Section \ref{sec:3}, a totally new stellar model of KIC 11145123 has been presented via the non-standard scheme based on the assumption that 
%the star was born as a single star with the ordinary initial helium abundance and, during the evolution, experienced some chemical composition modifications in the envelope 
%thought be caused by, for instance, mass accretion from the outside. 
%This is the first case where such non-standard modeling has been successfully carried out for the star. 
% (and based on the obtained envelope-modified model, we infer the internal rotation in the following section \ref{4-3}), 
A few discussions about the non-standard model of the star are given in this section, namely, 
the evolutionary history (Section \ref{sec:4-1}), 
the structure in the deep radiative region (Section \ref{sec:4-2}), 
and a possible relation between the internal structure and dynamics (Section \ref{sec:4-3}). 
% to the envelope-modified model are discussed 
%in this subsection focusing on the possibility of further improvements on the non-standard model of the star. 

\subsection{Single-star evolution or not?} \label{sec:4-1}
One of the goals in this paper is to constrain the evolutionary history of KIC 11145123, 
which is here discussed from two different perspectives, namely, in terms of stellar modeling (this study) 
and in terms of abundance analyses \citep{Takada_Hidai2017}. 

Though the envelope-modified models and the previous models are comparable with respect to 
the modeled frequencies (see discussions in Subsection \ref{sec:3-3-2}), 
we would like to emphasize that the envelope-modified model is representing 
KIC 11145123 better. 
This is because %the envelope-modified model can naturally explain the cause of the increase in 
we need not assume high initial helium abundance for the envelope-modified model. 
%the helium abundance in the envelope. 
In contrast, it is difficult to explain the origin of the deduced high initial helium abundance of the previous models 
constructed assuming single-star evolution; 
we have to accept complex scenarios such as that the star was born in some high-helium environments, 
which is actually proposed for some globular clusters, though the cause of which 
has not been understood yet as well \citep{Bastian2018}. 

Then, if we persist to a single-star scenario for the star, the star should be a peculiar single star 
that has been kicked out of a globular cluster. 
Although it is possible that the star has been kicked out of a globular cluster 
judging from the kinematics of the star \citep{Takada_Hidai2017}, 
the chance that the star is the kind of peculiar single stars with high helium abundance is small 
because the sodium enrichment and the carbon depletion, both of which are always confirmed for the kind of stars \citep{Bastian2018}, 
are not confirmed in the case of KIC 11145123 \citep{Takada_Hidai2017}. 
Combined with the fact that the abundance pattern of the star resembles those of typical blue straggler stars, 
it is probably the case that KIC 11145123 has experienced some interactions with other stars during the evolution. 

%As already mentioned in Subsection \ref{sec:3-2-3}, though the envelope-modified model reproduces 
%the observed g-mode period spacing patterns better than the previous models of e.g. \citet{Kurtz2014}, 
%there is 
Note that the envelope-modified model does not tell us about 
the corresponding progenitor because %of the assumption that $\delta m=0$ in our envelope-modifying computations 
$\delta M=0$ in our envelope-modifying computations; 
there are in principle an infinite number of possible progenitors with different masses 
from which the envelope-modified model can be computed 
by ``correctly'' modifying the progenitors with the corresponding $\delta M$ 
(see the latter note on this issue in the beginning of Section \ref{sec:2}). 
We nevertheless emphasize that 
%Important point is that 
what we originally would like to know was not the exact amount of mass gained by the star (via binary interactions or stellar merger) 
but whether the star could have experienced such envelope modification during the evolution or not, 
the latter of which can be achieved despite the fixed total mass $\delta M=0$. 
%even if we accept the assumption of the fixed mass coordinate $\delta m=0$.}

One mystery remains to be solved; the star is currently not in a binary system \citep{Takada_Hidai2017}, 
and how the envelope of the star has been modified is still unknown. 
Since \citet{Takada_Hidai2017} have utilized the so-called phase modulation analysis \citep{Murphy2016} 
to exclude the possibility of a binary system, 
they suggested that the star could belong to a binary whose spin-orbit is misaligned $90$ degrees (Murphy, 2018, private communication). 
Another possibility is that the star is a product of a more violent event such as stellar merger or stellar collision. 
This scenario naturally explains how the star has become a blue straggler star as well as 
the current rotational profile of the star that the envelope is rotating slightly faster than the deep radiative region \citep{Kurtz2014}, 
though it then challenges the origin of the very slow rotation of the star $P_{\mathrm{rot}}\sim100\,  \mathrm{d}$, i.e. 
a significant fraction of the angular momentum must have been lost from the star. % via, for example, 
Some numerical simulations have shown that the magnetic braking can 
contribute to the spin-down of collision products \citep[e.g.][]{Schneider2019}. 
However, \citet{Takada_Hidai2017} have failed to detect the magnetic field ($> 1 \, \mathrm{kG}$) for the star, 
further complicating the discussion about the origin of the star. 
To reveal the exact cause of the envelope modification that the star has experienced is, 
as noted in the previous paragraph, 
beyond the scope in this paper, but that is definitely an interesting subject to investigate in the future. %, 
%which might also help us to better understand, for instance, how the internal rotational profile of the star 
%that the envelope is rotating slightly faster than the deep radiative region has been realized. 

\subsection{G-mode period spacing ($\Delta P_{g}$) pattern revisited} \label{sec:4-2}
The second discussion is about the structure in the deep radiative region of the star. 
We especially concentrate on the chemical composition gradient left behind the receding nuclear burning core, 
which sensitively affects $\Delta P_{\mathrm{g}}$ patterns \citep[e.g.][]{Miglio2008}. 
After we show what impact steepness of chemical composition gradients has 
on the corresponding $\Delta P_{\mathrm{g}}$ patterns (Subsection \ref{sec:4-2-1}), 
we propose a possible resolution for a deviation between the modeled $\Delta P_{\mathrm{g}}$ pattern and the observation 
based on a realistic stellar model (Subsection \ref{sec:4-2-2}). 

%\subsubsection{Realizing steeper chemical composition gradients} \label{sec:4-2-1}
\subsubsection{Chemical composition gradients vs $\Delta P_{\mathrm{g}}$ patterns} \label{sec:4-2-1}
\begin{figure} [t]
 \begin{center}
  \includegraphics[scale=0.32,angle=270]{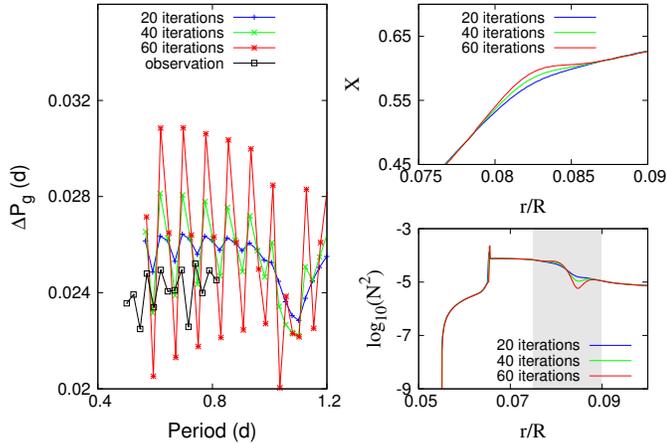}
  \caption{\footnotesize Numerically computed $\Delta P_{\mathrm{g}}$ patterns 
  (colored curves) and the observed $\Delta P_{\mathrm{g}}$ pattern (black) are shown in the left panel. 
  The corresponding internal structures, namely, the hydrogen profiles and 
  the $\rm{Brunt}$-$\rm{{V\ddot{a}is\ddot{a}l\ddot{a}}}$ frequencies 
  are illustrated in the right upper panel and the right lower panel, respectively. 
  Note that the grey-shaded area in the lower right panel corresponds to the range of the abscissa for the upper right panel. 
  %ranges of the abscissas are slightly different, which is indicated by a shaded area in the lower right panel. 
  The degree of perturbation (see equation (\ref{Eq_delta_mu_Gauss}) for one perturbation) 
  becomes larger in the order of blue, green, and red, and correspondingly, 
  an amplitude of an oscillatory component of a numerically computed $\Delta P_{\mathrm{g}}$ pattern becomes larger. 
  We can confirm that the chemical composition gradient is successfully steepened with our scheme (see the hydrogen profiles or the 
  $\rm{Brunt}$-$\rm{{V\ddot{a}is\ddot{a}l\ddot{a}}}$ frequencies at $r/R \sim 0.08$). } 
  \label{dPgs_ptbN_Ch3}
 \end{center} 
\end{figure}
%The second example is related to oscillatory behaviors of the $\Delta P_{\mathrm{g}}$ patterns deeply discussed in Chapter \ref{2}. 
%We here revisit the observed $\Delta P_{\mathrm{g}}$ pattern of KIC 11145123. 
%As it is demonstrated in Subsection \ref{2-4-4}, 
As demonstrated in Subsection \ref{sec:3-3-2}, the envelope-modified model successfully 
reproduces the positive slope of the observed $\Delta P_{\mathrm{g}}$ pattern (Figure \ref{dPg_N_tentative1}). 
Still, there is a discrepancy between the modeled $\Delta P_{\mathrm{g}}$ pattern 
(red in Figure \ref{dPg_N_tentative1}) and the observed one (black in Figure \ref{dPg_N_tentative1}) especially in terms of 
a short periodic component ($\Delta n \sim$ a few) seen in the observed $\Delta P_{\mathrm{g}}$ pattern. 

It is generally considered that $\Delta P_{\mathrm{g}}$ patterns can be analytically described as 
oscillating components (in terms of the g-mode radial order) whose periods and amplitudes are 
determined by locations and strengths, respectively, of the sharp features in the chemical composition gradients 
\citep[more precisely, the sharp features in the $\rm{Brunt}$-$\rm{{V\ddot{a}is\ddot{a}l\ddot{a}}}$ frequencies, 
see more discussions in, e.g.,][]{Miglio2008}. 

Thus, one possible approach to reproduce the shorter-period component of the observed $\Delta P_{\mathrm{g}}$ pattern is to 
add artificial perturbations in the chemical composition gradients 
and to render the gradients be steeper, %the $\rm{Brunt}$-$\rm{{V\ddot{a}is\ddot{a}l\ddot{a}}}$ frequency $\delta N^2$ (see Figure \ref{delta_N}). 
%However, the perturbation $\delta N^2$ is not physically motivated one and we do not know whether such structures are really feasible inside stars or not. 
%There is thus room for us to consider more physically motivated perturbations to the $\rm{Brunt}$-$\rm{{V\ddot{a}is\ddot{a}l\ddot{a}}}$ frequency, 
which can be achieved by applying the scheme %a function equipped in the scheme 
of constructing stellar models whose chemical compositions are arbitrarily modified (demonstrated in Section \ref{sec:2}). 
%as shown in the following paragraphs. 

Then, which part of the chemical composition gradient should we perturb? 
The period of the oscillatory component in the $\Delta P_{\mathrm{g}}$ pattern, in particular, is determined by 
the ratio $\Pi_{\mu}/\Pi_{0}$, where 
\begin{equation}
\Pi_{0}^{-1} = \int_{r_{0}}^{r_{1}} N d \, \mathrm{ln} \,r, \label{Eq_pi0} 
\end{equation} 
and 
\begin{equation}
\Pi_{\mu}^{-1} = \int_{r_{0}}^{r_{\mu}} N d \, \mathrm{ln} \,r \label{Eq_pimu} 
\end{equation} 
\citep{Miglio2008}. 
The $\rm{Brunt}$-$\rm{{V\ddot{a}is\ddot{a}l\ddot{a}}}$ frequency is denoted by $N$. 
The inner edge of the (high-order) g-mode cavity, the outer one, and the location of the sharp feature in $N$ are 
$r_{0}$, $r_{1}$, and $r_{\mu}$, respectively. 
%a ratio between the integration of the $\rm{Brunt}$-$\rm{{V\ddot{a}is\ddot{a}l\ddot{a}}}$ frequency throughout the star and 
%that 
%\edit1{Remembering that the ratio $\Pi_{\mu}/\Pi_{0}$ determines the oscillation period of the 
%g-mode period spacing, }
We thus perturb the outer region of the chemical composition gradient (see around 
$r/R\sim0.08$ in Figure \ref{dPgs_ptbN_Ch3}) 
with which the ratio $\Pi_{\mu}/\Pi_{0} \sim 3$, close to the short period ($\Delta n \sim$ a few) 
in the observed $\Delta P_{\mathrm{g}}$ pattern. 
%because it is theoretically expected that the period of the 
%oscillatory component in a $\Delta P_{\mathrm{g}}$ pattern is 
%
%It is generally considered that periods and amplitudes of $\Delta P_{\mathrm{g}}$ patterns are 
%determined by strengths and locations of the sharp features, respectively, in the chemical composition gradients 
%\citep[more precisely, the sharp features in the $\rm{Brunt}$-$\rm{{V\ddot{a}is\ddot{a}l\ddot{a}}}$ frequencies, 
%see more discussions in, e.g.,][]{Miglio2008}. 

Figure \ref{dPgs_ptbN_Ch3} illustrates $\Delta P_{\mathrm{g}}$ patterns which are numerically computed based on the 
perturbed %$\rm{Brunt}$-$\rm{{V\ddot{a}is\ddot{a}l\ddot{a}}}$ frequencies of the stellar 
models obtained using the resettling function; 
they are in both hydrostatic and thermal equilibrium states. 
In order to obtain the perturbed models, we perturb the deep radiative region, 
instead of the envelope as in Section \ref{sec:3}, so that 
the chemical composition gradient becomes steeper, 
using the following expression for one perturbation: 
\begin{equation}
\delta X \equiv  B_{\mathrm{G}} \times \mathrm{exp} \biggl [  -\frac{1}{2} \biggl ( \frac{m-m_{\mathrm{c}}}{A_{\mathrm{G}}} \biggr )^2  \biggr ], \label{Eq_delta_mu_Gauss} 
\end{equation} 
where $B_{\mathrm{G}}$, $m_{\mathrm{c}}$, and $A_{\mathrm{G}}$ are the amplitude, the center, and the 
width of the modification, respectively. 
The specific values are as follows: $B_{\mathrm{G}}=5\times 10^{-4}$, $m_{\mathrm{c}}=0.2826$, and 
$A_{\mathrm{G}}=10^{-2}$. 
%(for reasons why we focus on the chemical composition gradients, see discussions in Chapter \ref{2}), 
Then, the deviated models are resettled to hydrostatic states with our scheme (see details in Section \ref{sec:2}). 

It is clearly seen that our scheme successfully provides us with 
the equilibrium stellar models with the chemical composition gradients 
much steeper than that of the unperturbed model. 
It is also verified that an amplitude of the shorter-period component 
of a numerically computed $\Delta P_{\mathrm{g}}$ pattern becomes 
larger as the model is perturbed more (from blue to red in Figure \ref{dPgs_ptbN_Ch3}), 
which is the same trend as expected by \citet{Miglio2008}. 
%As such, our scheme (especially the resettling function) is also of great use for purposes 
%other than carrying out the non-standard modeling of stars. 

\subsubsection{Realizing steeper chemical composition gradients with a more realistic stellar model} \label{sec:4-2-2}
\begin{figure} [t]
\begin{center}
\includegraphics[scale=0.32,angle=270]{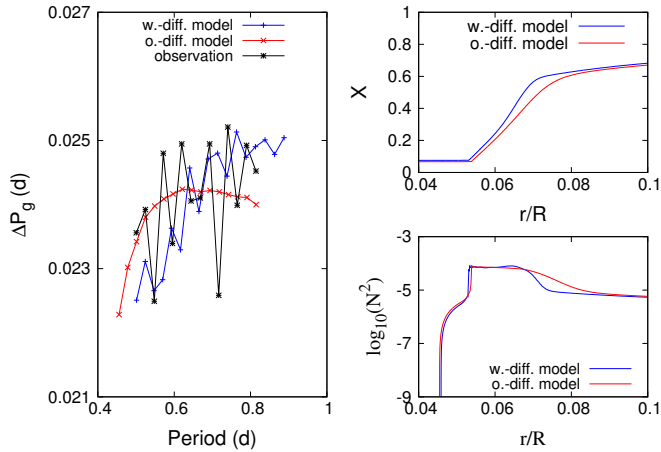}
\caption{\footnotesize Observed g-mode period spacings $\Delta P_{\rm{g}}$ (black thick line) 
and modeled g-mode period spacings (the red line and the blue line) in the left panel. 
The red line is calculated based on the envelope-modified model (here, denoted as ``o.-diff. model'', short for ``ordinary-diffusion model''), 
and the blue line is 
based on another model with weaker elemental diffusion (here, denoted as ``w.-diff. model'', short for ``weaker-diffusion model''). 
The latter model successfully reproduces the period of the 
short-periodic oscillatory component of the observed $\Delta P_{\rm{g}}$ pattern. 
We also see some improvements in terms of the amplitude of the short-periodic oscillatory component of 
$\Delta P_{\rm{g}}$ pattern compared with that of the envelope-modified model. 
The right panels show the corresponding hydrogen profiles (top) 
and $\rm{Brunt}$-$\rm{{V\ddot{a}is\ddot{a}l\ddot{a}}}$ frequencies (bottom). 
It is obvious that the chemical composition gradient of the weaker-diffusion model is 
much steeper than that of the envelope-modified model. }
\label{g_mode_period_spacing_now}
\end{center} 
\end{figure}
%This discrepancy has been already pointed out in Subsections \ref{2-4-3} and \ref{2-4-4}, and 
In the previous subsection, it has been suggested that 
to consider an artificial perturbation to the chemical composition gradient 
%$\rm{Brunt}$-$\rm{{V\ddot{a}is\ddot{a}l\ddot{a}}}$ frequency $\delta N^2$ 
could be helpful for reproducing the short-periodic component 
based on the direct numerical computations of the eigenfrequencies for several perturbed chemical composition gradients. 
%$\rm{Brunt}$-$\rm{{V\ddot{a}is\ddot{a}l\ddot{a}}}$ frequency. 
In particular, perturbing chemical composition gradients so that the gradients become steeper 
is a possible solution to reproduce the 
shorter-period component of the observed $\Delta P_{\mathrm{g}}$ pattern. % as shown in Section \ref{3-6}. 
Then, the next question is what is the mechanism that is at work during the evolution and produces such structures?

%In this subsection, elemental diffusion process much weaker than usually expected 
%is assumed to be playing a key role in reproducing a chemical composition gradient 
%which leads to the short-periodic component in the observed $\Delta P_{\mathrm{g}}$ pattern. 
%This is due to the fact that elemental diffusion renders the chemical composition gradient less steeper. 
%The validity of the assumption is partly confirmed by stellar equilibrium models computed without diffusion processes during evolution (see Subsection \ref{2-3-2} and Figure \ref{dPgs_N_1p4M_2M_diff_vs_ovs}); the $\Delta P_{\mathrm{g}}$ patterns calculated based on the models 
%exhibit short periodic component with high amplitudes. 
%Actually, the amplitudes are too large to represent the observed pattern for KIC 11145123. 
One of the straightforward ways to produce such steep chemical composition gradients 
is not to activate elemental diffusion in 1-d stellar evolutionary calculations; 
the diffusion process renders the chemical composition gradient less steep. %steeper. 
Still, elemental diffusion is generally expected to be at work inside stars \citep{Michaud2015}, 
and thus, in this subsection, we consider a model with elemental diffusion, 
but ``much weaker" diffusion processes during evolution. 

We implement such ``much weaker" diffusion by changing the default criterion 
for the maximum diffusion velocity ($\mathrm{diffusion} \_ \mathrm{v} \_ \mathrm{max}$ $=$ $1.$d$-$3 adopted in MESA) 
to a much smaller value ($\mathrm{diffusion} \_ \mathrm{v} \_ \mathrm{max}$ $=$ $1.$d$-$10). 
Specifically, we force the gravitational settling to be suppressed (note that we do not include radiative levitation 
in our computations as described in Section \ref{sec:2.5}). 

In MESA, the diffusion velocity of each element in each mass shell is 
computed by solving Burgers' equation \citep{Burger1969}, which sometimes leads to 
unphysically large diffusion velocities in, for instance, the outermost envelope. 
The criterion $\mathrm{diffusion} \_ \mathrm{v} \_ \mathrm{max}$ in MESA is thus usually set to avoid such problems, and 
our implementation is rather crude in a sense that our purpose of setting $\mathrm{diffusion} \_ \mathrm{v} \_ \mathrm{max}$ 
is different from the original one as \citet{Paxton2011} have proposed. 
Nevertheless, we clearly see improvements in the behavior 
of a short periodic component in the $\Delta P_{\mathrm{g}}$ pattern 
computed with the much weaker diffusion processes during evolution 
(see Figure \ref{g_mode_period_spacing_now}), showing a high potential of the implementation 
for further asteroseismic researches. 

It should be noted, however, that the implementation is 
computationally fairly time-consuming, and that it is still hard 
to incorporate the scheme into, for example, the grid-based modeling of stars. 
This is due to the presence of discontinuities in chemical composition gradients 
caused by the weaker diffusion, leading to very small timesteps in 
stellar evolutionary computations \citep[see more details in][]{Paxton2011}. 

Another point worth mentioning is that radiative levitation, which 
has been phenomenologically treated with the scheme of \citet{Morel2002} in this study, 
can counteract the gravitational settling to 
reduce net velocities for elemental diffusion as shown by \citet{Deal2018}, 
and thus, including radiative levitation in computations of elemental diffusion 
could be a promising next step to be done in the near future. 

\subsection{A relation between the inferred steeper $\nabla_{\mu}$ and the fast-core rotation?} \label{sec:4-3}
In the previous subsection, we see that adopting ``much weaker'' diffusion in our computations partly successfully reproduces 
the observed $\Delta P_{\mathrm{g}}$ pattern. 
But we cannot immediately conclude that the diffusion process inside the star is really weak 
since mixing processes other than convection, convective overshoot, and diffusion are 
neglected in our 1-d stellar evolutionary computations, among which 
rotation is known to cause extra mixing processes inside stars as well as to 
counteract diffusion processes \citep[e.g.][]{Deal2020}. 
%The point is that the hydrogen abundance (or the $\rm{Brunt}$-$\rm{{V\ddot{a}is\ddot{a}l\ddot{a}}}$ frequency) profile of 
%the weaker-diffusion model represents that of the star better than that of the envelope-modified model, 
%and thus, we have to reconsider mixing processes whichever result in the inferred hydrogen abundance profile.} 

%It is generally considered that a velocity shear causes instabilities 
%around the shear boundary and finally mixes the region to some extent. 
%Actually, 
Interestingly, \citet{Hatta2019} have pointed out the possibility that the convective core of KIC 11145123 
is rotating 5--6 times faster than the other regions of the star. 
Therefore, for KIC 11145123, it is expected that the inferred fast-core rotation, 
or the inferred rotational velocity shear between the convective core and the radiative region above, 
can cause instabilities which lead to mixing around the convective core boundary. 

Then, let us focus on such extra mixing caused by rotational shear instabilities. 
We have the following two possibilities regarding the inferred steeper chemical composition gradient of the star. 
The first possibility is that diffusion process is really weak (more precisely, somehow much weaker than the current 
theoretical computations predict) and the rotationally induced mixing can be ignored (though the rotational velocity 
shear does exist), leading to the inferred chemical composition gradient. 
Or, the second possibility is that the diffusion is working as the theory predicts but 
the rotationally induced mixing effectively counteracts the diffusion, leading to the inferred chemical composition gradient. 
%Because the theoretical framework of diffusion computations have been fairly well-established \citep{Burger1969}, 
%we consider the second possibility more probable in this study. 
%Although it is almost impossible to directly observe such ``extra'' mixing which is at work deep inside the star, 
%we can find a signature of the extra mixing 
%based on the discussions about the observed $\Delta P_{\mathrm{g}}$ pattern (Section \ref{sec:4-2}); 
%it is necessary for us to somehow weaken the diffusion process inside the star 
%to reproduce the observed $\Delta P_{\mathrm{g}}$ pattern, 
%and it is possible that the extra mixing 
%caused by the rotational velocity shear 
%around the convective boundary counteracts and effectively weakens the diffusion processes. 
%We thus find a hint for a relation between the internal dynamics and the structure of the star.} 

Actually, a relation, similar to that suggested by the second possibility, 
between internal dynamics and structure can be also found in the case of the Sun. 
According to results of helioseismic structure inversion, 
there is a discrepancy between the sound speed profile of the real Sun and 
that of the standard solar model at the bottom of the solar convective envelope \citep{Christensen-Dalsgaard1996}. 
Interestingly, the bottom of the convective envelope 
where the discrepancy has been found is close to the so-called solar tachochline, 
a relatively strong rotational velocity shear inferred based on helioseismic rotation inversion \citep{Thompson1996}, 
and it is currently commonly accepted in the helioseismology community 
that the discrepancy can be resolved if we consider extra mixing 
caused by the velocity shear at the solar tachochline. 
\citep[See more detailed discussions in, e.g.,][]{Gough1996,JCD2021}. 

The second possibility that the rotation-induced mixing counteracts the elemental diffusion 
thus seems to be more attractive %probable 
compared with the first possibility that the elemental diffusion is really weak. 
But here is one caveat; in helioseismology, extra mixings at the solar tachochline are 
believed to mix the region uniformly (reducing effectiveness of the helium gravitational settling), but 
in the case of KIC 11145123, we have an opposite trend 
where extra mixings around the convective boundary render the chemical composition gradient to be steeper somewhere. 
The latter process might sound peculiar for us because mixing processes, literally, mix the chemical composition uniformly. 
However, whether a mixing process really leads to a locally uniform chemical composition profile or not 
strongly depends on the scaleheight of the mixing process and the position where the mixing is at work; 
if the mixing region is too thin and the mixing is occurring 
at the edge of the boundary (in terms of the chemical composition profile, for instance), 
the gradient of the chemical composition profile can be maintained, or even strengthen, 
though the chemical composition is uniform inside the thin mixing region, which might be the case for KIC 11145123. 
\begin{figure} [t]
 \begin{center}
  \includegraphics[scale=0.32,angle=270]{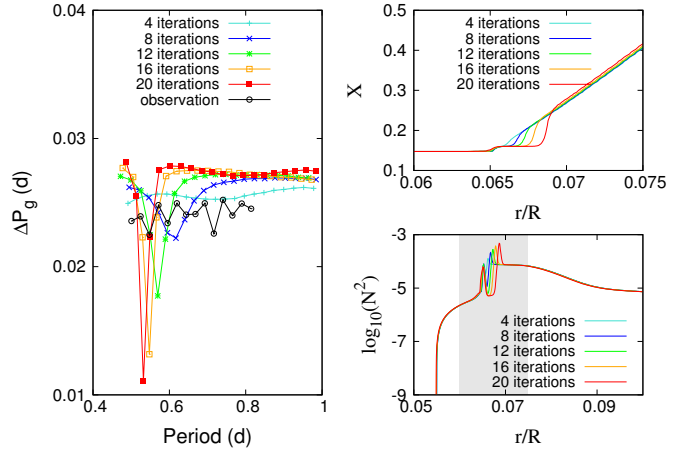}
  \caption{\footnotesize Numerically computed $\Delta P_{\mathrm{g}}$ patterns (left), 
  hydrogen mass contents (upper right), and $\rm{Brunt}$-$\rm{{V\ddot{a}is\ddot{a}l\ddot{a}}}$ frequencies (lower right) 
  of models whose chemical composition gradients are modified. 
  The modifications are given mimicking extra mixings 
  caused by rotational shear instabilities around the convective boundary, leading to chemically uniform regions 
  such as those seen between $0.065 < r/R < 0.07$ (see the upper right panel). 
  Note that the chemical compositions in the modified region are not exactly the same as those in the convective core 
  due to the thermally-evolving step in the resettling scheme (see the last paragraph of Section \ref{sec:2-1}). 
  It is seen that as we render the fully mixed region broader (from light blue to red), 
  the dip in the $\Delta P_{\mathrm{g}}$ patterns become larger (see also light blue curves to red curves). 
  The numerically computed $\Delta P_{\mathrm{g}}$ patterns 
  nevertheless do not reproduce a shorter-period component 
  as that confirmed in the observed $\Delta P_{\mathrm{g}}$ pattern (black). 
  The grey-shaded area in the lower right panel corresponds to the range of the abscissa for the upper right panel, 
  as in Figure \ref{dPgs_ptbN_Ch3}. } 
  \label{dPgs_ptbN_rot_sh_expand}
 \end{center} 
\end{figure}

To test the possibilities, we again utilize the resettling function to obtain stellar models 
whose chemical composition gradients are modified, as demonstrated in Subsection \ref{sec:4-2-1}. 
In this case, we have mimicked the extra mixing originating from rotational shear instabilities 
by artificially rendering the chemical composition gradients developed 
just above the convective core to be uniform; in the modified region, 
the chemical compositions are homogeneous and almost the same as 
those in the convective core (Figure \ref{dPgs_ptbN_rot_sh_expand}). 
Once we resettle the perturbed model to the hydrostatic and thermal equilibrium state, 
we have extended the uniformly modified region by $6\times10^{-4}$ in fractional radius per one modification, and 
repeated the process described above until the upper boundary of the artificially well-mixed region reaches $r/R\sim0.068$ 
(which corresponds to the red curves in Figure \ref{dPgs_ptbN_rot_sh_expand}). 
%
%In this case, we first artificially render the chemical composition gradients developed 
%just above the convective core to be uniform  
%(mimicking the extra mixing caused by rotational shear instabilities), 
%in order to evaluate the effects of the 

Figure \ref{dPgs_ptbN_rot_sh_expand} shows structures of thus obtained models (right panels) and 
the corresponding $\Delta P_{\mathrm{g}}$ patterns (left panel). 
It is clearly seen that the chemical composition gradients can be locally steeper in the presence of the 
relatively small-scale mixing, which has been expected from the discussions 
in the preceding paragraphs. 
However, we do not see any improvements in the numerically computed $\Delta P_{\mathrm{g}}$ patterns; 
the observed short-period oscillatory component is not reproduced by any modified models. 
This is because, in this case, the ratio $\Pi_{\mu}/\Pi_{0}$ (see the expressions \ref{Eq_pi0} 
and \ref{Eq_pimu}) %, where the position of the sharp feature in the 
%$\rm{Brunt}$-$\rm{{V\ddot{a}is\ddot{a}l\ddot{a}}}$ frequency corresponds to that of the 
%artificially generated steeper chemical composition gradient, 
is much larger than $3$, which is expected to reproduce the shorter periodic component in the observed $\Delta P_{\mathrm{g}}$ pattern, 
resulting in a longer periodic component in the numerically computed $\Delta P_{\mathrm{g}}$ patterns 
(see discussions in Section \ref{sec:4-2}). 
We therefore cannot explain the observed $\Delta P_{\mathrm{g}}$ pattern based on the 
assumption that the rotationally induced mixing is counteracting the elemental diffusion process, 
and we consider the first possibility, that the diffusion process is somehow really weak in the 
deep region of the star, more probable compared with the other possibility. 

%% Putting eqnarrays or equations inside the mathletters environment groups
%% the enclosed equations by letter. For instance, the eqnarray below, instead
%% of being numbered, say, (4) and (5), would be numbered (4a) and (4b).
%% LaTeX the paper and look at the output to see the results.

\section{Conclusions} \label{sec:5}
This paper is dedicated to detailed asteroseismic non-standard modeling of a possible blue straggler star, KIC 11145123. 
Two main conclusions have been obtained and they are listed in the following paragraphs. 

The first conclusion is that the star might have been born as a single star 
with an ordinary initial helium abundance of $\sim 0.26$ and then 
experienced some interactions with other stars, 
leading to the modification of the chemical composition in the envelope. 
This conclusion is obtained based on the non-standard asteroseismic modeling of the star 
where modifications of the chemical compositions 
are taken into account for constructing 1-dimensional stellar models. 
A scheme to compute such non-standard models 
has been developed in this study, which is applied to the comprehensive grid-based modeling of the star, 
resulting in the envelope-modified model with fundamental parameters as below: 
$M=1.36M_{\odot}$, $Y_{\mathrm{init}}=0.26$, 
$Z_{\mathrm{init}}=0.002$, and $f_{\mathrm{ovs}}=0.027$. %, and $ \mathrm{Age}=2.169 \times 10^{9}$ years old. 
The modification is down to the depth of $r/R\sim0.6$ 
and the extent is $\Delta X \sim 0.06$ ($\Delta X$ is the difference in hydrogen abundance 
between the unmodified model and the modified model) at the surface. 
This is the first time such an envelope-modified model 
(which is still in both hydrostatic and thermal equilibrium states) is obtained for the star, 
and the discrepancy between the modeled eigenfrequencies and the observed ones is comparable 
to those for previous models computed based on an assumption of a single-star evolution 
and high initial helium abundance. %, leading to the first conclusion. 
The conclusion that this star may well have experienced some interactions with other stars 
during the evolution is consistent with the formation channels of blue straggler stars, 
thus strengthening the argument that the star is a (probable rather than possible) blue straggler star. 
%\edit1{In addition, combined with the fact that the chance of the star currently having a companion is fairly small, 
%and that the envelope of the star is rotating slightly faster than the deep radiative region, 
%the star might have experienced rather violent events such as stellar collisions or mergers, 
%subsequently lost a significant fraction of its angular momentum via, e.g., the magnetic braking 
%%(or the total angular momentum ), 
%and finally resulted in the very slow rotation period $P_{\mathrm{rot}}\sim100\, \mathrm{d}$. }
%\edit1{To investigate the exact cause of the envelope modification of the star 
%would be a next subject worth dealing with, which might help us better understand, for instance, 
%}

The second conclusion is that the elemental diffusion in the deep region of the star 
might be much weaker than that assumed in ordinary stellar evolutionary calculations. 
%
%re might be an extra mixing 
%caused by the rotational velocity shear 
%which are thought to be at work around the convective core boundary of the star. 
%Though exact mechanisms are not clear yet, 
This conclusion is obtained based on the detailed analysis 
of the observed $\Delta P_{\mathrm{g}}$ pattern of the star, which evidently suggests that 
the chemical composition gradient in the deep radiative region above the convective core 
should be much steeper than that of the envelope-modified model. 
%Though we do not see a clear relation between the previously inferred rotational velocity shear and 
%the inferred steeper chemical composition gradient for the star, 
%It is shown that extra mixings caused by the rotational velocity shears which has been inferred by previous studies can 
%maintain the chemical composition gradient steeper, but 
%there should be some sharp features in the Brunt-$\rm{V}\ddot{a}is\ddot{a}l\ddot{a}$ frequency 
%and that 
%\edit1{the extra mixing caused by the rotational velocity shear effectively counteracts the elemental diffusion process, 
%which is represented by the fact that }
%``much weaker" diffusion than 
%that adopted in ordinary 1-dimensional stellar evolutionary codes such as MESA is favorable to the observation. 
%\section{Software and third party data repository citations} \label{sec:cite}
Though exact mechanisms to render the chemical composition gradient so steep are not clear yet, 
this is the first study which tests the possibility 
that there is a relationship between the current rotational profile and the structure of the stars, and 
to investigate such relationship should be one of the most highly prioritized subjects to future researches. 

\acknowledgments

We would like to thank the NASA and \textit{Kepler} team for the data of inestimable value. 
We thank D. W. Kurtz for his constructive comments. 
We are grateful to M. Takada-Hidai for his insightful suggestions. 
H. Saio is also thanked for his advices regarding the non-standard modeling. 
Y.H. acknowledges the Research Fellowship from the Japan Society for the Promotion 
of Science for Young Scientist.

\end{document}